\documentclass[backref, breaklinks, twocolumn, colorlinks]{aastex63}
\hypersetup{linkcolor=blue,citecolor=blue,filecolor=cyan,urlcolor=magenta}

\usepackage{enumitem}
\usepackage{mathtools}
\usepackage{amssymb}
\usepackage{amsmath}
\usepackage{float}
\usepackage{ulem}
\usepackage{placeins}
\usepackage{epstopdf}
\usepackage{color}
\usepackage{graphicx,epsfig}
\usepackage{xcolor}
\usepackage{graphicx}
\usepackage{wasysym}

\usepackage{wasysym,MnSymbol}

\newcommand{\nicer}{{\sl NICER }}
\newcommand{\mypsr}{PSR~J0030+0451 }
\newcommand{\mypsR}{PSR~J0030+0451}

\def\gsim{\mathrel{\raise.5ex\hbox{$>$}\mkern-14mu
             \lower0.6ex\hbox{$\sim$}}}

\def\lsim{\mathrel{\raise.3ex\hbox{$<$}\mkern-14mu
             \lower0.6ex\hbox{$\sim$}}}

\shorttitle{Multipoles in MSPs}

\begin{document}

\title{{The Multipolar Magnetic Field of Millisecond Pulsar PSR J0030+0451}}

\correspondingauthor{Constantinos Kalapotharakos}
\email{constantinos.kalapotharakos@nasa.gov
\\ ckalapotharakos@gmail.com}

\author[0000-0003-1080-5286]{Constantinos Kalapotharakos}
\affiliation{University of Maryland, College Park (UMCP/CRESST II)\\
College Park, MD 20742, USA} \affiliation{Astrophysics Science
Division, NASA/Goddard Space Flight Center\\ Greenbelt, MD 20771,
USA}

\author[0000-0002-9249-0515]{Zorawar Wadiasingh}
 \affiliation{Astrophysics Science Division,
NASA/Goddard Space Flight Center\\ Greenbelt, MD 20771, USA}
\affiliation{Universities Space Research Association (USRA)\\
Columbia, MD 21046, USA}

\author[0000-0001-6119-859X]{Alice K. Harding}
\affiliation{Astrophysics Science Division, NASA/Goddard Space
Flight Center\\ Greenbelt, MD 20771, USA}

\author[0000-0002-7435-7809]{Demosthenes Kazanas} \affiliation{Astrophysics Science
Division, NASA/Goddard Space Flight Center\\ Greenbelt, MD 20771,
USA}

\begin{abstract}
{Modeling of the \nicer X-ray waveform of the pulsar \mypsR, aimed to
constrain the neutron star mass and radius, has inferred surface
hot-spots (the magnetic polar caps) that imply significantly
non-dipolar magnetic fields. To this end, we investigate magnetic
field configurations that comprise offset dipole plus quadrupole
components using static vacuum field and force-free global
magnetosphere models. Taking into account the compactness and
observer angle values provided by \citet{Miller2019} and
\citet{Riley2019}, we compute geodesics from the observer plane to
the polar caps to compute the resulting X-ray light curve. We
explore, through Markov chain Monte Carlo techniques, the detailed
magnetic field configurations that can reproduce the observed X-ray
light curve and have discovered degeneracies, i.e., diverse field
configurations, which can provide sufficient descriptions to the
\nicer X-ray waveforms. Having obtained the force-free field
structures, we then compute the corresponding synchronous 
$\gamma$-ray light curves following \cite{2014ApJ...793...97K}; 
these we compare to those obtained by \emph{Fermi-LAT}, to provide models 
consistent with both the X-ray and the $\gamma$-ray data, thereby restricting
further the multipole field parameters. An essential aspect of this
approach is the proper computation of the relative phase between the
synchronous X- and $\gamma$-ray light curves. We conclude with a discussion of
the broader implications of our study.}
\end{abstract}
\vspace{-0.1in}

\section{Introduction}\label{intro}

Rotation-powered pulsars are known to exhibit emission across the
entire observed electromagnetic spectrum, from radio to TeV
$\gamma$-rays. While not all energetic pulsars have been detected in
all bands, it is generally accepted that this is mainly a result of
the observers' orientation to the neutron star (NS) spin-axis. The
radio and thermal X-rays in pulsars are thought to be emitted at low
altitudes (the X-rays on the surface and the radio not very far from
it), the result of dissipation of rotational energy into electric
currents and particle acceleration. Because of its location, the
X-ray emission has been considered as a reliable probe of NS masses
$M_{\star}$ and radii $r_{\star}$, through the detailed modeling of
the pulsar X-ray light curves, including the effects of photon
propagation in the non-Minkowski metric of the NS
{\citep{2013ApJ...776...19L,2015ApJ...808...31M}}. The precise
determination of $\{M_{\star}, r_{\star}\}$ has been one of the
goals of the \nicer \citep{2016SPIE.9905E..1HG} mission. To this
end, one of the first \nicer targets has been the millisecond pulsar
(MSP) \mypsr (with a spin period of $4.865$ ms), which is also a
radio and a $\gamma$-ray pulsar, the first $\gamma$-ray MSP
announced by \emph{Fermi-LAT} \citep{Abdo2009}.

\defcitealias{Miller2019}{M19}
\defcitealias{Riley2019}{R19}

Recently, two groups, \citet{Miller2019}, \citet{Riley2019},
hereafter \citetalias{Miller2019} and \citetalias{Riley2019},
respectively, reported strong evidence for the existence of
multipolar magnetic fields through modeling of the X-ray waveforms
of PSR J0030+0451 that constrain $\{M_{\star}, r_{\star}\}$ to
unprecedented accuracy. {These results require the soft X-ray
emitting areas on the NS surface to be both located in one
rotational hemisphere while the observer direction lies in the other
hemisphere\footnote{{In this study, without loss of generality, we
always consider that the observer is in the Northern rotational
hemisphere (i.e., the hemisphere that is defined by
$\pmb{\Omega}\cdot \hat{\mathbf{n}}>0$, where $\pmb{\Omega}$) is the
spin angular frequency vector and $\hat{\mathbf{n}}$ the normal
vector at any point of the surface of the NS.}}, with one spot being
compact and the other a more elongated oval or crescent}. Since
these spots are thought to be generated by energetic particles from
polar-cap pair-cascades \citep{Harding2001} or return currents
\citep{1999ApJ...511..351C,2010ApJ...715.1282B,2014ApJ...793...97K},
such a configuration implies that the NS magnetic field cannot be a
centered dipole, which would produce two near-circular antipodal
polar caps \citep{2009ApJ...699.1171A,Bilous2019}.

\citet{2015MNRAS.450..714P,2016MNRAS.463.1240P}
provided exact vacuum solutions of retarded fields 
for multipolar and offset dipole magnetic fields, respectively.
{\citet{2017ApJ...851..137G} presented semi-analytic field
structures that incorporate a superposition of centered dipole and
quadrupole magnetic moments that are co-aligned. This configuration
supports circular and annular polar caps the centers of which are
antipodal. The imposed symmetries do not seems to be consistent with
the results of \citetalias{Miller2019,Riley2019} for \mypsR.}

Over the last decade, global macroscopic
\citep{2010MNRAS.404..767C,2010ApJ...715.1282B,2014ApJ...793...97K}
and kinetic particle-in-cell (PIC)
\citep{2016MNRAS.457.2401C,Philippov2018,2018ApJ...857...44K} pulsar
magnetosphere models have determined that the main site of particle
acceleration and MeV-GeV $\gamma$-ray emission is the equatorial
current sheet (ECS) (although some emission can occur at high
altitudes inside the light-cylinder (LC)), a fact supported by the
recent discovery of the $\gamma$-ray pulsar Fundamental Plane
\citep{2019ApJ...883L...4K,2020arXiv200810821P}. The ECS is a
characteristic feature of the force-free (FF) solutions
\citep{1999ApJ...511..351C,2006MNRAS.368.1055T,2006ApJ...648L..51S,2009A&A...496..495K}
and forms outside the LC radius $R_{\rm LC}=c/\Omega$.

The immediate question then is whether such a distorted, far from a
pure centered dipolar, field geometry on the NS surface would
manifest in the pulsar $\gamma$-ray light curves. While the location
of the high-energy emission is generally considered remote (i.e.,
distant) from the NS surface, where the \emph{NICER}-implied
magnetic field distortions are important, the LC of \mypsr is only
$R_{\rm LC} \approx 18 r_{\star}$ (assuming the \nicer inferred
radius of $r_{\star} \approx 13$ km). To address this issue,
\citet{Chen2020} presented recently a NS centered vacuum
dipole-plus-offset-quadrupole magnetic field configuration, whose
footpoints on the NS surface roughly match the emission regions of
\citetalias{Riley2019}. Adjusting the parameters of this general
field make-up, they constructed a global FF magnetosphere model that
produced $\gamma$-ray, radio, and X-ray light curves in {modest}
agreement with observations. Their approach was driven mainly by the
requirement that the centered dipolar component reproduce the
$\gamma$-ray light curve. This led them to impose a magnetic dipole
inclination angle of $80^\circ$.

In this paper, we take a different and more rigorous
approach: We assume that the magnetic field comprises off-center
dipole and quadrupole components and determine their parameters
(i.e. positions, directions, relative strength) under static vacuum
conditions, by demanding that the corresponding polar caps reproduce
accurately the \nicer X-ray light curve. This we achieve by
employing a newly developed code, {\tt{GIKS}}, that ray-traces
photons in the Kerr-metric from a distant observer to the stellar
surface hot-spots, for each choice of the magnetic field structure
parameters. We also explore the corresponding parameter space of
field configurations through a Markov Chain Monte Carlo (MCMC) code
we have developed. We then turn the field configuration from static
vacuum field (SVF) to force-free (FF) and implement the MCMC code to
modify appropriately the field parameters so that the X-ray light
curves of the FF configurations remain consistent with observation;
for each such choice we also compute the corresponding $\gamma$-ray
light curves and through their comparison with the \emph{Fermi-LAT} ones we
finally obtain an optimum set of magnetic field parameters
consistent with both the X-ray and the $\gamma$-ray light curves. A
significant nuance we consider is the {proper phase synchronization
of X-rays and $\gamma$-rays, which requires accounting for light
travel time effects between the X-ray and the $\gamma$-ray emission
regions.}

In \S\ref{metho}, we describe our codes and methodology and in
\S\ref{results} our results for the SVF and FF field configurations.
Finally, in \S\ref{conclusions}, we summarize the main
conclusions, and in \S\ref{discussion}, we discuss the broader
impacts and validity of our study as well as the future prospects.

\section{Methodology}\label{metho}

\subsection{\textbf{G}eodesic \textbf{I}ntegration in \textbf{K}err \textbf{S}pacetime (\tt{GIKS})}\label{codeRT}

Our ray-tracing code integrates an ensemble of individual photon
trajectories that are placed on a uniform grid on a predefined
sufficiently distant image (i.e., observer) plane with initial
photon velocities perpendicular to that plane \citep[similar to, but
an independent implementation of, those detailed
in][]{2010ApJ...718..446J,2012ApJ...745....1P,2012ApJ...753..175B}.
This approach requires a much smaller number of integrations for a
given density of trajectories that reach the image plane compared to
an approach that integrates trajectories originating on the NS
surface. We adopt a full Kerr-metric\footnote{{We note that even
though {\tt GIKS} code incorporates the Kerr spacetime the
corresponding frame-dragging effects in this type of problem are
actually negligible, which implies that our results are practically
consistent with Schwarzschild spacetime. A more rigorous approach
would require the implementation of a spacetime (i.e., metric) that
is the result of the numerical solution of the corresponding full
Einstein equations with a specified equation of state.}} but
currently neglect stellar oblateness, which is expected to have
{smaller} effects for the rotational frequency value of
\mypsR\footnote{{The ellipticity for the rotational frequency of
\mypsr is expected to be $\lesssim 6\times 10^{-3}$
\citep{2013ApJ...777...68B}.}}. The impact positions and zenith
angles, $\vartheta_{\rm z}$, of the photons that reach the stellar
surface are stored. The trajectories that hit the polar caps, which
are assumed to be the hot-spot regions, are identified, taking into
account the stellar rotation corresponding to the travel time of the
individual photons. Doppler-boosting and emission anisotropy are
incorporated by adopting photon weights on the image plane
\citep{2014ApJ...792...87P}. The light-curve intensity at a specific
phase (i.e., observer time) is then derived by integrating over the
image plane. The process is repeated for different phases that
sample a complete period. The results are binned in phase
identically to the binning of the observed X-ray light curve. We
have verified our independent code on several test cases presented
in \cite{2014ApJ...792...87P, 2019ApJ...887L..26B}.

In all the calculations below, we always use libraries of $\approx
2.5\times 10^6$ photon trajectories that originate from an observer
plane that is located at a distance of 300 Schwarzchild radii (i.e.,
$r_{\rm obs}=300r_{\rm s}$, where $r_{\rm s}=2GM_{\star}/c^2$) and
reach the stellar surface\footnote{A value for the observer distance
10 times higher {does not affect the results}.}. We have verified
the numerical fidelity of the model X-ray light curves due to the
adopted $r_{\rm obs}$ value and the finite number of photon
trajectories and it is well below the observational uncertainties
for J0030+0451.

{For our models of the \mypsr X-ray profile, we fix $M_{\star}$,
$r_{\star}$, and observer viewing angle, $\zeta$, from the results
of either \citetalias{Miller2019}\footnote{We considered the two
oval model. Nonetheless, we note that \citetalias{Miller2019} also
presented a model with three hot-spots in which the smallest
hot-spot had a different {(i.e., larger)} temperature than the other
two {but due to its location and size had minor contribution}. This
model was {only marginally} statistically preferred compared to the
one with the two equal-temperature oval shaped hot-spots.} or
\citetalias{Riley2019}}\footnote{We considered the best model (i.e.,
ST+PST).}. Moreover, \citetalias{Miller2019} and
\citetalias{Riley2019} reported that the temperatures of the two
hot-spots are almost identical and so we assumed that all the
hot-spots have the same temperature. Although the \nicer inference
codes used a fully ionized H atmosphere model, we currently
implement, for simplicity, a $\cos^b{\vartheta_{\rm z}}$
distribution for the emission from the hot-spots, where
$\vartheta_{\rm z}$ is the zenith angle and $b$ the assumed
anisotropy index. Exponents in the range $b \sim 0.5 - 1.0$ are a
good approximation to the NSX atmosphere used by \nicer (W. Ho,
private Comm.). {Testing different $b$ values, for the hot-spot
shapes corresponding to the highest likelihood parameter sets of
\citetalias{Miller2019} and \citetalias{Riley2019}, we found that
the \mypsr X-ray profile is reproduced for $b=0.65$ and $b=1.0$,
respectively.} We note that \citetalias{Miller2019} used channels 40
- 299 ($0.4 - 3$ keV) while \citetalias{Riley2019} included \nicer
energy channels 25 - 299 ($0.25 - 3$ keV) in their analyses.

\begin{figure*}[t]
\vspace{0.0in}
  \begin{center}
    \includegraphics[width=0.99\textwidth]{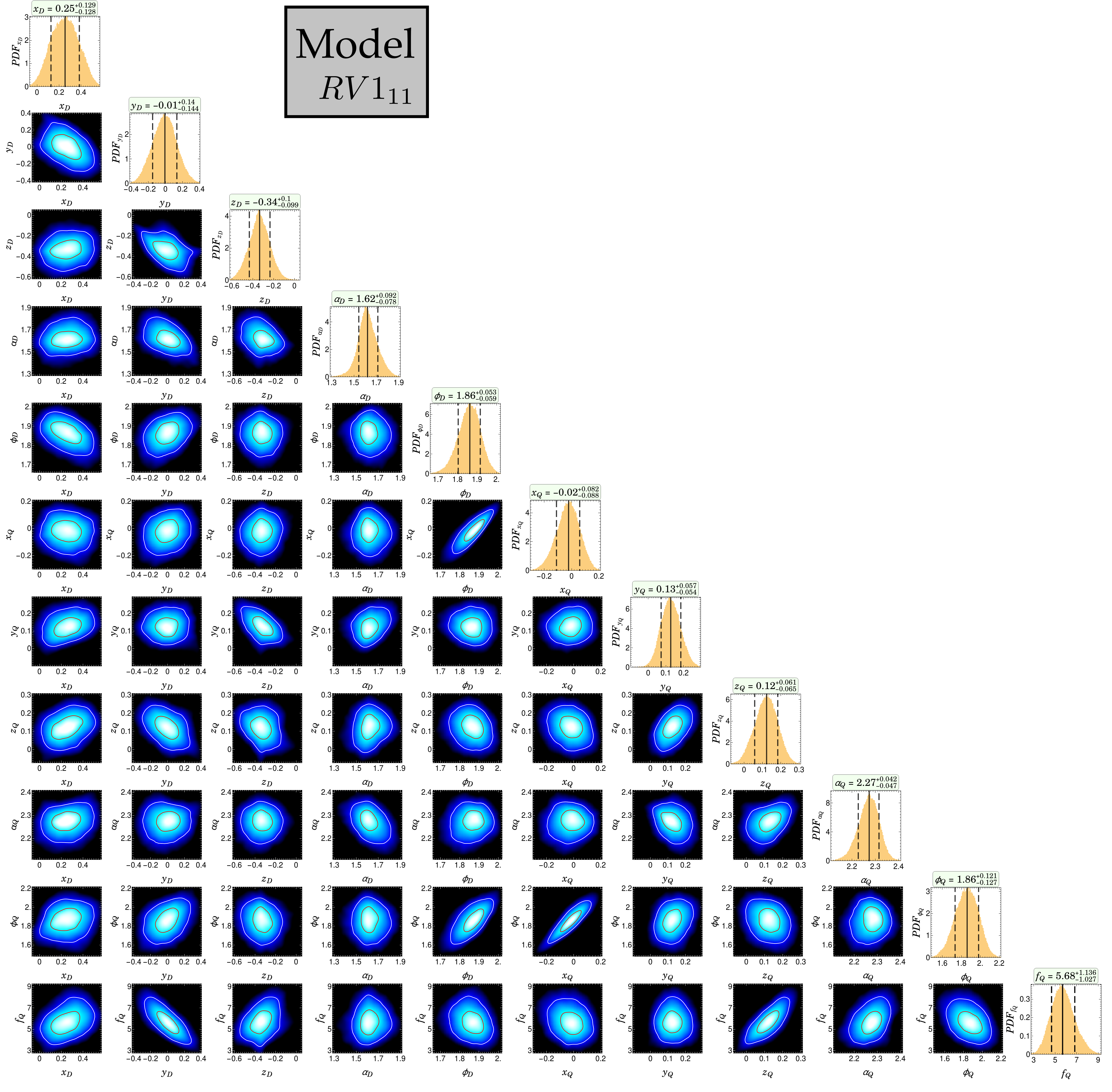}
  \end{center}
  \vspace{-0.3in}
  \caption{The corner plot, produced by the MCMC exploration
  corresponding to the model $RV1_{11}$. All the possible
   2D posterior probability density distributions are shown in
  density plots from low (black) to high (white) values. The two lines in
  each panel denote the $1\sigma$ and $2\sigma$ regions. The
  marginalized 1D posterior probability distributions for the
  11 parameters are shown along the diagonal. The solid lines
  denote the corresponding mean values while the dashed lines
  denote the $68\%$ containment regions. The median and $\pm~\sigma$
  values for each parameter are shown above the corresponding
  1D plot.}
  \label{fig:01}
  \vspace{0.0in}
\end{figure*}

\begin{deluxetable*}{C|||CCCCCC|CCCCCC}[t]
\tablenum{1} \tablecaption{Parameters corresponding to the highest
likelihood for the SVF models. For the $RV$ models, we adopted the
\citetalias{Riley2019} {median values (of the corresponding
posterior distribution)} for the stellar mass
$M_{\star}=1.34M_{\bigodot}$, radius $r_{\star}=12.71$km, and the
observer angle $\zeta=53.85^{\circ}$. Similarly, for the $MV$
models, we adopted the \citetalias{Miller2019} {best fit} values
$M_{\star}=1.49M_{\bigodot}, r_{\star}=13.64$km, and
$\zeta=47.38^{\circ}$. We emphasize that the compactness
$(M_{\star}/r_{\star})$ between the $RV$ and $MV$ models are much
closer to each other compared to the corresponding individual
$M_{\star}$ and $r_{\star}$ values. \label{tab:vac}}
\tablewidth{0pt} \tablehead{ \cline{2-13} \colhead{Quantity} &
\colhead{$RV1_{11}$} & \colhead{$RV2_{11}$} & \colhead{$RV3_{11}$} &
\colhead{$RV4_{11}$} & \colhead{$RV5_{8}$} & \colhead{$RV6_6$}  &
\colhead{$MV1_{11}$} & \colhead{$MV2_{11}$} & \colhead{$MV3_{11}$} &
\colhead{$MV4_{11}$} & \colhead{$MV5_{8}$} & \colhead{$MV6_6$} }
\startdata
$x_D (r_{\star})$        &    0.12 &    0.17 &    0.45 &   -0.20 &    0.00 &    0.00 &    0.26 &    0.26 &    0.07 &    0.34 &    0.00 &    0.00 \\
$y_D (r_{\star})$        &   -0.17 &    0.39 &    0.29 &    0.46 &    0.00 &    0.00 &    0.12 &    0.22 &   -0.36 &   -0.21 &    0.00 &    0.00 \\
$z_D (r_{\star})$        &   -0.30 &   -0.04 &    0.11 &   -0.25 &    0.00 &    0.00 &    0.24 &   -0.35 &   -0.24 &   -0.40 &    0.00 &    0.00 \\
$\alpha_D  ({\rm rad})$   &    1.75 &    1.60 &    1.98 &    1.53 &    1.00 &    1.38 &    1.38 &    1.72 &    1.29 &    1.45 &    1.40 &    1.35 \\
$\varphi_D  ({\rm rad})$  &    1.94 &    2.19 &    2.21 &    4.60 &    1.82 &    1.94 &    2.56 &    2.56 &    5.79 &    2.60 &    2.63 &    2.53 \\
 \cline{1-1}
$x_Q (r_{\star})$        &    0.07 &    0.40 &    0.16 &    0.31 &    0.62 &    0.00 &    0.09 &    0.06 &   -0.23 &   -0.36 &    0.09 &    0.00 \\
$y_Q (r_{\star})$        &    0.14 &    0.22 &   -0.17 &   -0.13 &    0.20 &    0.00 &    0.16 &    0.17 &   -0.37 &    0.36 &    0.19 &    0.00 \\
$z_Q (r_{\star})$        &    0.07 &   -0.12 &   -0.38 &   -0.15 &   -0.20 &    0.02 &   -0.17 &   -0.18 &   -0.25 &   -0.09 &   -0.17 &   -0.16 \\
$\alpha_Q  ({\rm rad})$   &    2.18 &    2.29 &    0.60 &    1.68 &    0.92 &    0.75 &    2.34 &    2.18 &    2.04 &    2.60 &    2.38 &    2.42 \\
$\varphi_Q  ({\rm rad})$  &    1.90 &    2.45 &    6.06 &    4.94 &    5.78 &    4.89 &    2.69 &    2.72 &    0.07 &    2.17 &    2.65 &    2.35 \\
$B_{\rm Q}/B_{\rm D}$    &    5.58 &    5.01 &    2.39 &    3.22 &    8.69 &    6.38 &    4.97 &    3.58 &    8.10 &    0.93 &    4.81 &    5.17 \\
 \hline
$\chi_{\rm r}^2$       &    0.66 &    0.60 &    0.71 &    0.78 &    0.58 &    0.85 &    0.70 &    0.72 &    1.36 &    1.67 &    0.68 &    0.80 \\
\enddata
\tablecomments{Prefixes: R -- Riley, M -- Miller, V -- static vacuum
fields} \vspace{0.0in}
\end{deluxetable*}

\subsection{Markov Chain Monte Carlo Exploration}
\label{codeMCMC}

We have developed an MCMC code\footnote{A custom {\tt{Fortran}}
{code with serial} and MPI-parallel implementations of a stretch
move of \cite{2010CAMCS...5...65G}.} that computes likelihoods for
the walker steps corresponding to different field parameters (i.e.,
different field structures and therefore, different hot-spots) for
the \nicer bolometric (0.25 - 1.45 keV) X-ray profile from
\citet{Bogdanov2019}. We defer the full energy-dependent analysis
for a future work.

The adopted field model configuration is the superposition of static
dipole and $m=0$ quadrupole moments whose orientation and offsets
(from the center of the NS) are allowed to vary. {Flat distributions
are considered for all the priors while the {radial} offsets are
limited to not exceed $0.7r_{\star}$.} The dipole and quadrupole
fields, in spherical magnetic coordinates, read
\begin{align}\label{bdbqfields}
    \mathbf{B_{\rm D}}&=B_{\rm
    D}\left(\frac{r_{\star}}{r_{\rm D}}\right)^3
    \left\{\cos\theta_{\rm D},\;\;\frac{1}{2}\sin\theta_{\rm D},\;\;0
    \right\}\\
    \mathbf{B_{\rm Q}}&=B_{\rm
    Q}\left(\frac{r_{\star}}{r_{\rm Q}}\right)^4
    \left\{\frac{1}{2}(3\cos^2\theta_{\rm Q}-1),\;\;\cos\theta_{\rm Q}\;\sin\theta_{\rm Q},\;\;0
    \right\}
\end{align}
where $r_{\rm D},~r_{\rm Q}$ are the corresponding radial distances
from the moment locations and $\theta_{\rm D},~\theta_{\rm Q}$ the
corresponding magnetic co-latitudes.

The determination of the magnetic field configuration requires ten
parameters. More specifically, two for the position of the dipole
moment relative to the star center (at some fixed azimuth), three
for the position of the quadrupole moment relative to the dipole
moment, four angles that determine the orientation of the dipole and
quadrupole moments with respect to $\pmb{\Omega}$, and the
$\pmb{\Omega}$-observer plane, and the relative strength of the
quadrupole moment (i.e., $f_{\rm Q}\equiv B_{\rm Q}/B_{\rm D}$) at a
distance that is equal to the NS radius. Thus, without regard for
absolute phasing, ten parameters describe the fields. However, for a
finite distance between the star and the image plane, the absolute
azimuth of the dipole moment is also set by the presumed zero phase
point (e.g. from radio) of the X-ray light curve.

\begin{figure*}[th]
\vspace{0.0in}
  \begin{center}
    \includegraphics[width=1.0\textwidth]{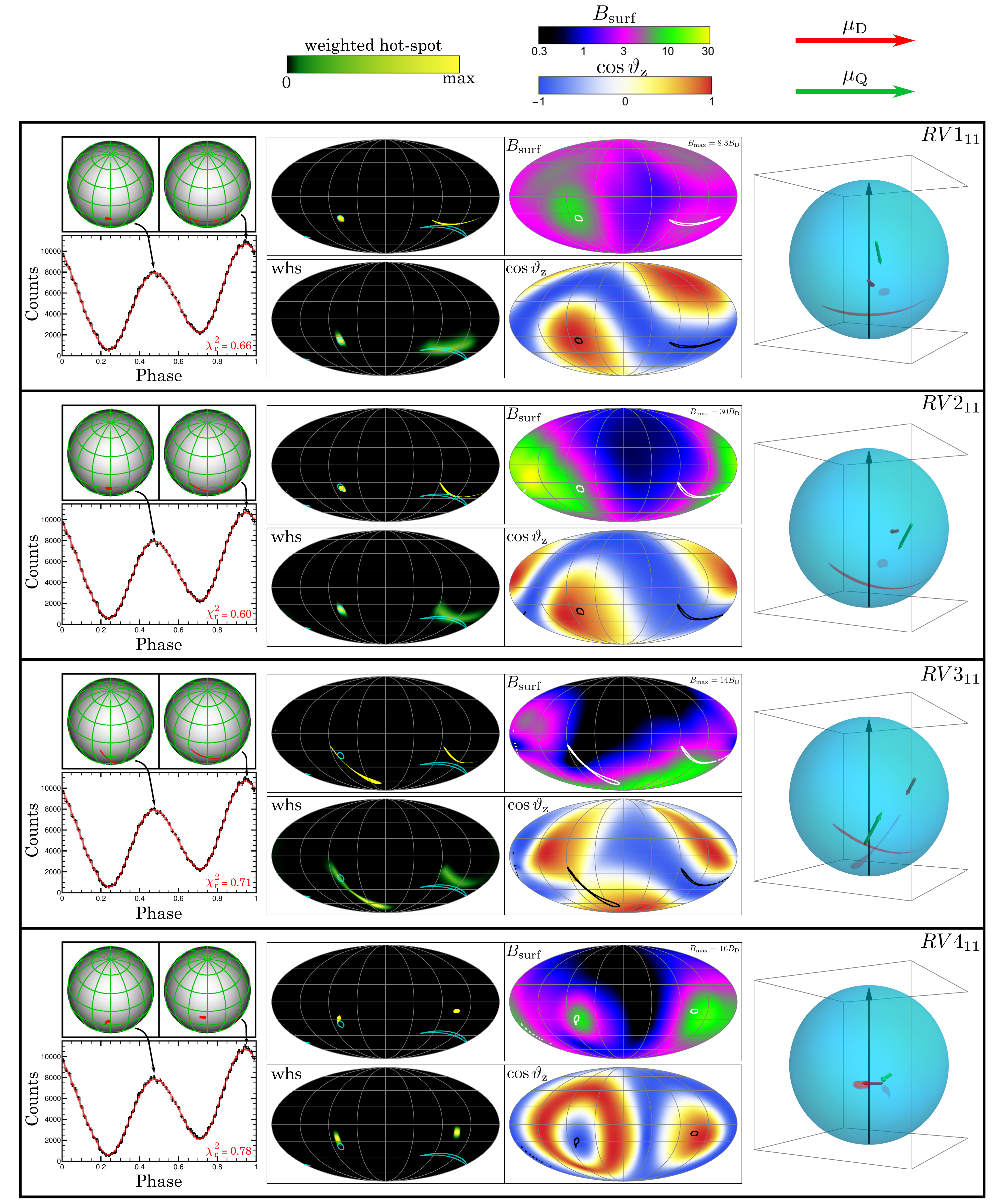}
  \end{center}
  \vspace{-0.23in}
  \caption{The characteristics of the four $RV$ models with eleven free
  parameters. Each row (inside the thick black frame) corresponds to
  the indicated model. In the first column, the bottom panel shows the
  observed \nicer X-ray light curve (black line with error-bars)
  together with the corresponding model X-ray light curve (red color).
  The corresponding $\chi^2_{\rm r}$ values are depicted in the panels.
  (Continued in the next page)}
  \label{fig:02}
  \vspace{0.0in}
\end{figure*}

\addtocounter{figure}{-1}
\begin{figure*}[ht]
\vspace{0.0in}
  \begin{center}
  \end{center}
  \vspace{-0.23in}
  \caption{(Continued) The two top panels depict the image plane, with limb darkening, at the indicated phases (i.e., the primary and secondary maxima). The top panels in the
  second column depict the model hot-spots (i.e., polar caps)
  in yellow color on the Mollweide projection of the stellar
  surface. The bottom panels show, in the indicated color scale, the superposition
  of hot-spot regions from the entire corresponding posterior distribution
  (i.e., weighted hot-spots). In both panels, the cyan lines denote
  the hot-spot regions denoted by \citetalias{Riley2019}. The top panels
  of the third column depict, in the indicated color scale,
  the strength of the surface magnetic field. The corresponding maximum
  strength is also indicated. The bottom panels depict, in the indicated
  color scale, the cosine of the {zenith} angle of the magnetic field
  on the stellar surface. The model hot-spots are indicated in the top
  and bottom panels by the white and black lines, respectively.
  The fourth column schematically depicts (in flat spacetime) the locations of the dipole (red arrow)
  and the quadrupole (green arrow) magnetic moments inside the star.
  The locations of the moments are at the middle of
  the plotted arrows. The black arrows indicate $\pmb{\Omega}$.
  {The corresponding model polar caps (red areas) are also depicted
  on the transparent stellar surface.}}
  \label{fig:02}
  \vspace{0.0in}
\end{figure*}

\begin{figure*}[tbh]
\vspace{0.0in}
  \begin{center}
    \includegraphics[width=0.99\textwidth]{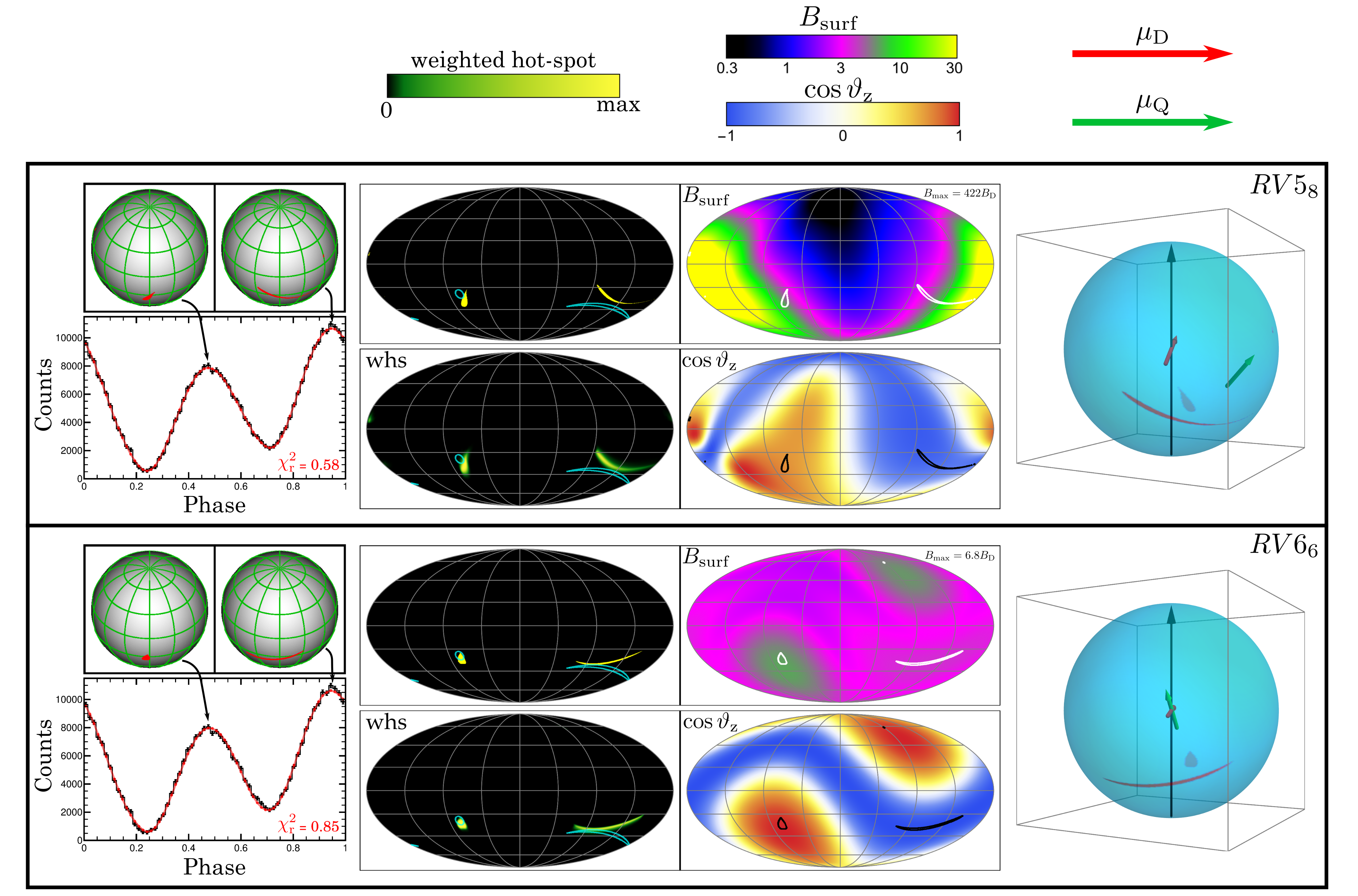}
  \end{center}
  \vspace{-0.25in}
  \caption{Similar to Fig.~\ref{fig:02} but for the $RV$ models with 8 and
  6 free parameters.}
  \label{fig:03}
  \vspace{0.0in}
\end{figure*}

Thus, eleven free parameters are explored by the MCMC, which are the
three Cartesian position coordinates $\{x_{\rm D},~y_{\rm D},~z_{\rm
D}\}$, the inclination $\alpha_{\rm D}$ and the azimuth direction
$\phi_{\rm D}$ of the dipole moment, the corresponding three
Cartesian position coordinates $\{x_{\rm Q},~y_{\rm Q},~z_{\rm
Q}\}$, the inclination $\alpha_{\rm Q}$ and the azimuth direction
$\phi_{\rm Q}$ of the quadrupole moment, and the $f_{\rm Q}$ ratio.
We note that the Cartesian coordinates are measured with respect to
the center of the star, the inclination angles with respect to
$\pmb{\Omega}$, and the azimuth direction angles with respect to the
$x$ axis oriented with $\pmb{\Omega}$. In this configuration the
observer always lies on the $x-z$ plane. The eleven parameters
described above define the magnetic field configuration (at the
assumed distance from the image plane) at the moment the observer
{receives} the zero-phase photons. To clarify, the ``picture'' the
observer observes at the time corresponding to the zero-phase photon
is different (due to the $r_{\rm obs}/c$ travel time) from what the
field parameter values dictate (see also Fig.~\ref{fig:08} and the
related discussion in Section~\ref{FFfldfermi}).

\begin{figure*}[!tbh]
\vspace{0.0in}
  \begin{center}
    \includegraphics[width=0.99\textwidth]{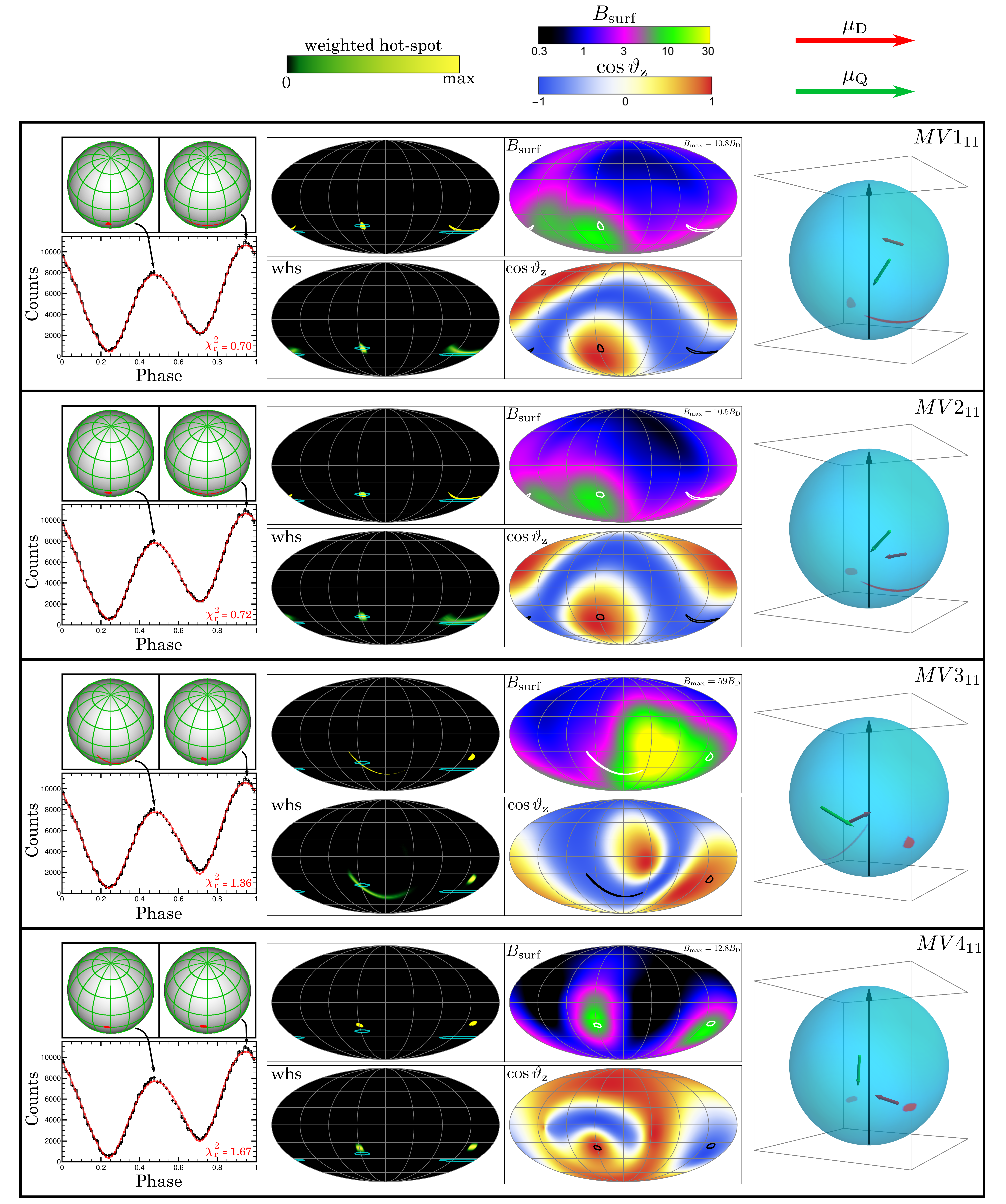}
  \end{center}
  \vspace{-0.2in}
  \caption{Similar to Fig.~\ref{fig:02} but for the $MV$ models with 11
  free parameters. {The cyan lines in the second column denote the hot-spot
  regions denoted by \citetalias{Miller2019}. We note that the apparent
  phase difference between the \citetalias{Riley2019} hot-spots shown
  in Figs.~\ref{fig:02}--\ref{fig:03} and the \citetalias{Miller2019}
  ones shown in Figs.~\ref{fig:04}--\ref{fig:05} is due to the different
  adopted $r_{\rm obs}$ values (recall that $r_{\rm obs}=300r_{\rm s}$
  and therefore, different in the two cases). }}
  \label{fig:04}
  \vspace{0.0in}
\end{figure*}

\begin{figure*}[!tbh]
\vspace{0.0in}
  \begin{center}
    \includegraphics[width=0.99\textwidth]{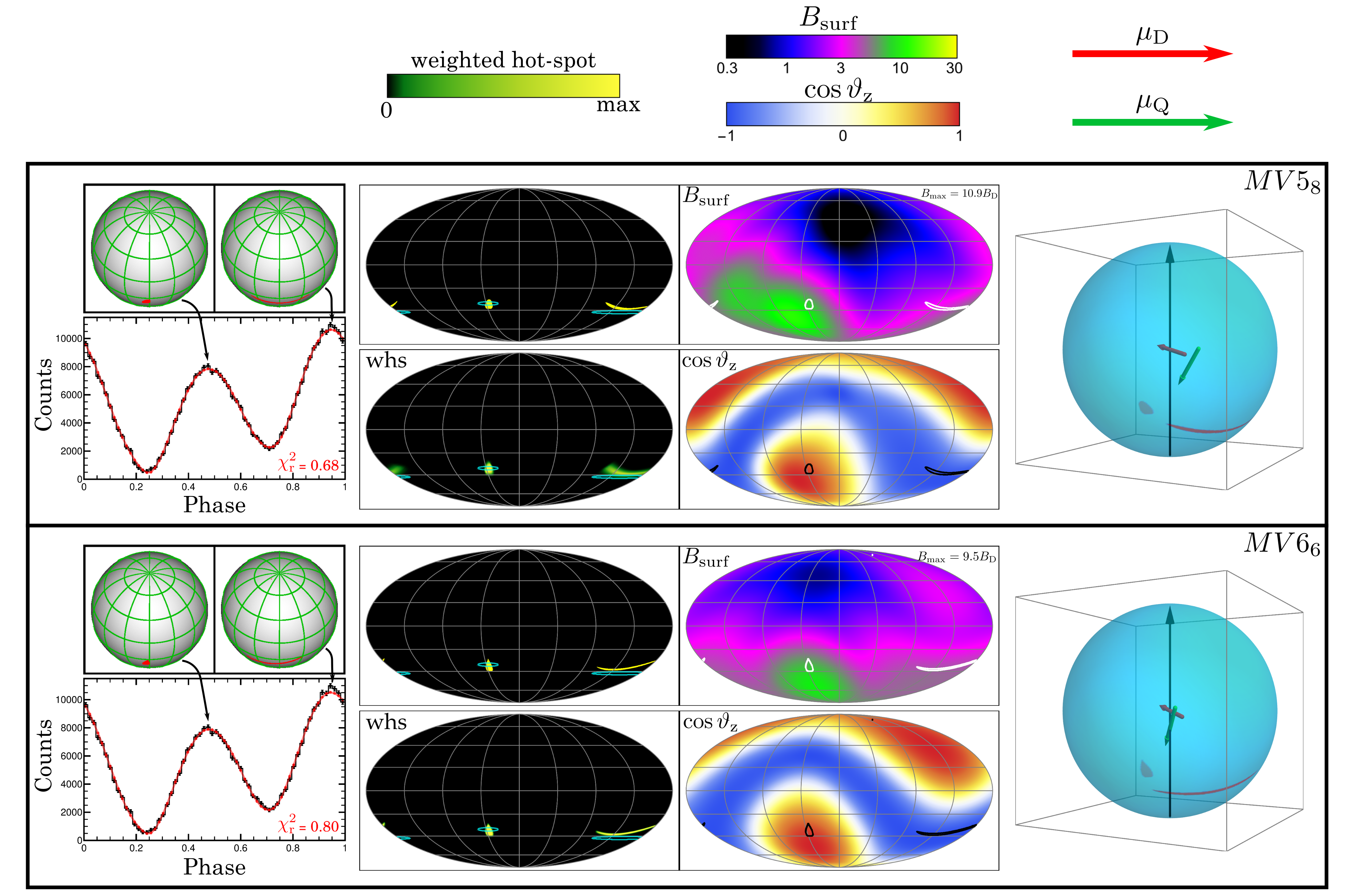}
  \end{center}
  \vspace{-0.2in}
  \caption{{Similar to Fig.~\ref{fig:04} but for the $MV$ models with 8 and
  6 free parameters.}}
  \label{fig:05}
  \vspace{0.0in}
\end{figure*}

In each MCMC step, the polar cap corresponding to the field
structure, determined by a point of the eleven dimensional parameter
space, is calculated, and then the model X-ray light curve is
derived using the library of photon trajectories that has been
produced by {\tt{GIKS}}. Comparison of the model to the observed
X-ray light curve then provided the likelihood value and the
corresponding $\chi^2$ value. We note that the adopted background
level of the \nicer X-ray light curve has been chosen to be
consistent with the background level depicted in
\citetalias{Miller2019}. Thus, the signal reads $N_{\rm S}=N_{\rm
T}-N_{\rm B}$, where $N_{\rm S},~N_{\rm T}$, and $N_{\rm B}$ are the
source, total, and background photon counts, respectively. The
adopted uncertainties for $N_{\rm T}$ and $N_{\rm B}$ are considered
Poisson type (i.e., $\sqrt{N_{i}}$, where $N_{i}$ is the number of
observed counts in the $i^{\rm th}$ bin out of the 64 phase bins).
Therefore, the uncertainty of the source photon count is
$\sigma_{\rm S}=\sqrt{\sigma_{\rm T}^2+\sigma_{\rm
B}^2}=\sqrt{N_{\rm T}+N_{\rm B}}$.

Finally, we note that assuming equal hot-spot temperatures, the
normalization of our model photon weights for the X-ray flux is
determined using the reported model hot-spot shapes from
\citetalias{Miller2019} and \citetalias{Riley2019}, which reproduce
the bolometric light curves for $b=0.65$ and $b=1.0$, respectively
(see end of section \ref{codeRT}).

\vspace{0.35in}
\section{Results}
\label{results}

\subsection{Model Nomenclature}
In this section, we present our results adopting the
findings of \citetalias{Riley2019} assuming both SVF and FF field
structures and the findings of \citetalias{Miller2019} assuming SVF
structures. We note that our models are named $ABn_m$, where $A=(R,
M)$ indicates the adopted fixed parameters (\citetalias{Riley2019}
or \citetalias{Miller2019}), $B=(V, F)$ indicates the assumed field
structure, static vacuum field (SVF) or force-free (FF), $n$
describes the number of the corresponding $AB$ model, and $m$
indicates the dimensionality of the corresponding field parameter
space.

\subsection{Vacuum Fields}
\label{vacfld}

In the first step, we explored the parameter space assuming that the
global magnetosphere structure is described by an SVF. Since the
SVFs are described by analytic expressions, the rapid calculation of
the polar caps allows both a test of our methodology and a more
thorough exploration of the parameter space on the order of seconds
per likelihood evaluation.

We separately ran 2000 independent serial MCMC chains each of which
started from different random initial positions, in the parameter
space.

The average number of accepted MCMC steps (according to the
corresponding likelihood ratios) in each of these chains was $\sim
800$. This exploration located distinct islands in the
multidimensional parameter space with high likelihood. We then
ranked the likelihood values of these distinct islands and selected
the four highest locales, which represent local maxima in a
multimodal landscape.

We then thoroughly explored the islands individually, using the
parallel version of the MCMC code. In each of these MCMC
explorations, the total number of accepted points, excluding the
burn-in period, were $\sim 4\times 10^5$.

Moreover, following a similar process, we studied models with
restricted priors. More specifically, we fixed the dipole moment at
the center of the NS and additionally also restricted the quadrupole
moment along the spin axis. In these cases, the dimensionality of
the parameter space is reduced from 11 to 8 and 6, respectively.

In Table~\ref{tab:vac}, we present the model parameters for a total
of twelve (six adopting the \citetalias{Riley2019} {median
$M_{\star},~r_{\star}$, and $\zeta$ values and six adopting the
\citetalias{Miller2019} best fit ones}) cases. The fifth and sixth
parameter sets (i.e., $RV5_{8},~MV5_{8},~RV6_{6},~MV6_{6}$) are the
highest likelihood models for the centered dipole plus free
quadrupole and for the centered dipole plus $z$-axis quadrupole.
{The last row shows the corresponding reduced $\chi^2$ values},
$\chi^2_{\rm r}={\chi^2}/{{\rm dof}}$ where dof are the degrees of
freedom, which in our case is ${\rm dof}=n-k$, where $n=64$ the
number of bins and $k$ the dimensionality of the parameter space.
{We note that the specific parameter values that are presented in
all the Tables have been calculated using a surface resolution
$5000\times 5000$ for the hot-spot determination while the surface
resolution that was used for the calculations that concluded to the
posterior distributions (e.g., Fig.~\ref{fig:01}) was $600 \times
600$. The determination of hot-spots is made by the identification
of the open magnetic field lines (i.e., those that cross the LC).
For the identification of the open magnetic field lines, we
integrate, using an adaptive Runge-Kutta 4--5 method, $5000\times
5000$ (or $600\times 600$) field lines uniformly distributed on the
stellar surface. {A bilinear interpolation is then used to describe
the entire stellar surface.} The errors in fidelity of the model
X-ray light curves due to the $600\times 600$ discretization are
well below the observed uncertainties (while for $5000\times 5000$
are negligible).} We also note that in the current study, we have
calculated the fields in the Euclidean space\footnote{{Assuming a
pure centered dipole, the field direction inside the corresponding
polar cap on the stellar surface of \mypsr is diverted by $\lesssim
0.5^{\circ}$ due to the corresponding Schwarzschild spacetime
distortions \citep{1983ApJ...265.1036W,1994ApJ...425..767G}. The
diversion of the field direction decreases almost inversely
proportional to the $r/r_{\star}$.}}.

All the models describe the bolometric \nicer X-ray light curve
extremely well. It is noted though that the $MV$ models seem to have
slightly higher $\chi_r^2$ values than those of the $RV$ ones.

In Fig.~\ref{fig:01}, we present the eleven parameter corner plot
for the $RV1_{11}$. The density-plots show, in color-scale, the
projected distributions on the corresponding 2D {subspaces}. The red
and white lines denote the $1\sigma$ and $2\sigma$ regions,
respectively. Along the diagonal the histograms of the marginalized
1D projections of the {11-dimensional} parameter-space are plotted.
The solid and dashed lines denote the median and $1\sigma$ values,
which are also shown above the histogram panels.

In Figs.~\ref{fig:02}--\ref{fig:03}, we present the characteristics
for all six $RV$ models from Table~\ref{tab:vac}. The first column
shows the model X-ray light curve (red line) together with the
observed one (black points and error-bars). The image planes
corresponding to the indicated light-curve peaks are also depicted.
The hot-spot (i.e., emitting) regions are the red areas. We note
that the gray color-scale denotes the adopted emission anisotropy.
In the top panel of the second column, the model hot-spots (yellow
areas) corresponding to the highest-likelihood parameter sets are
plotted on the stellar surface Mollweide projection. The areas
within the cyan lines denote the hot-spot areas of
\citetalias{Riley2019}. The bottom panel of the second column
presents, in the indicated color-scale, the weighted (according to
the corresponding posterior probability distributions) model
hot-spots. The third column shows properties of the magnetic field
structure on the projected stellar surface. The top and bottom
panels present, in the indicated color-scales, the magnitude
{$B_{\rm surf}$ and cosine of the zenith angle} (i.e., $\cos
\vartheta_{\rm z}$) of the magnetic field. The highest values of the
corresponding magnetic fields (in units of the centered dipole field
on the stellar surface) are indicated in the top panels. The white
and black lines in these plots denote the corresponding hot-spots.
The last column shows the locations and directions of the
corresponding dipole moment (red arrow) and quadrupole moment (green
arrow). Finally, Figs.~\ref{fig:04}--\ref{fig:05} are similar to
Figs.~\ref{fig:02}--\ref{fig:03} but for the $MV$ models (see
Table~\ref{tab:vac}).

Our results show a degeneracy of the field structures that can
reproduce the bolometric X-ray light curve of \mypsR. It is also
evident that quite different field parameters can provide similar
hot-spot regions on the stellar surface (see $[RV2_{11}, RV5_{8}]$,
$[RV1_{11}, RV6_{6}]$, and $[MV1_{11}, MV2_{11}, MV5_{8}, MV6_{6}]$
model family groups). Many hot-spots of our models share common
features to those presented by \citetalias{Miller2019} and
\citetalias{Riley2019}. Thus, the left-hand hot-spot is often more
{compact and} circular while the right-hand is elongated. However,
the elongated hot-spots in our models always form a ``smiley'' face
while \citetalias{Miller2019}'s and \citetalias{Riley2019}'s ones
form a neutral and ``frowny'' faces, respectively\footnote{{A
detailed exploration of the parameter space has shown that there are
regions of rather limited hypervolume in the parameter space that
allow wavy hot-spot shapes that contain a significant ``frowny''
component. Nonetheless, none of these can adequately describe the
bolometric X-ray light curve.}}. The difference of the hot-spot
shapes between \citetalias{Miller2019} and \citetalias{Riley2019}
was indicative of the underlying degeneracy as well as their
different adopted shape geometries. Moreover, there are cases in our
models where the hot-spots are quite dissimilar qualitatively than
those of \citetalias{Miller2019} and \citetalias{Riley2019}.

The maximum field value on the stellar surface in our models ranges
from $\sim 7$ to $\sim 400B_{\rm D}$. The quadrupole strength is
higher than the dipole one with only one exception (i.e.,
$MV4_{11}$). For eleven (out of the twelve) cases the $f_{\rm Q}$
ratio is $f_{\rm Q} \gtrsim 2.4$ while for the eight of them it is
$f_{\rm Q} \gtrsim 4.8$. The $f_{\rm Q}$ values corresponding to
parameter spaces with lower dimensionality (i.e.,
$RV5_{8}~RV6_{6}~MV5_{8}~MV6_{6}$) are always $f_{\rm Q} \gtrsim 5$.

\subsection{Force-Free Plasma Models}\label{FFfld}

\subsubsection{The \emph{NICER} X-ray Light Curve}\label{FFfldnicer}

The SVF {solutions are} instructive and helpful for code testing,
but FF and dissipative magnetosphere models with currents and
charges are much closer to those of real pulsars. Furthermore, such
models have the ability to show the distributions of currents of
different signs and values on the polar caps, which are physically
important for constraining the pair-cascades and heating of the
polar caps (i.e., hot-spots). Dissipative magnetospheres with
accelerating electric field components also enable us to compute
$\gamma$-ray emission light curves. Having determined the SVF
configuration that reproduces the \nicer X-ray profile, these
parameter values were used as initialization coordinates for the
serial MCMC to find the best parameter FF solution for a dipole plus
quadrupole field configuration, again fixing {the
$M_{\star},~r_{\star}$, and $\zeta$ values} from the \nicer results.

We have incorporated the FF magnetosphere simulator
\citep{Kala2012,2014ApJ...793...97K} inside our MCMC code as a
module for likelihood evaluation. The MCMC part of the code dictates
the magnetic field parameter values that are explored and the FF
simulator provides the corresponding magnetosphere structure. The
set of the field parameter values define the field boundary
conditions on the stellar surface. The magnetic field inside the
star that is determined in principle by the sum of the offset dipole
and quadrupole moments while the electric field is defined by
$\mathbf{E}=-(\pmb{\Omega}\times \pmb{r})\times\mathbf{B}/c$.

\begin{deluxetable*}
{C|||CCCCCC}[t] \tablenum{2} \tablecaption{Parameters corresponding
to the highest likelihood for the FF models. \label{tab:ff}}
\tablewidth{0pt} \tablehead{ \cline{2-7} \colhead{Quantity} &
\colhead{$RF1_{11}$} & \colhead{$RF2_{11}$} & \colhead{$RF3_{11}$} &
\colhead{$RF4_{11}$} & \colhead{$RF5_{8}$} & \colhead{$RF6_6$} }
\startdata
$x_D (r_{\star})$        &    0.16 &    0.26 &    0.48 &   -0.13 &    0.00 &    0.00 \\
$y_D (r_{\star})$        &   -0.21 &    0.29 &    0.23 &    0.60 &    0.00 &    0.00 \\
$z_D (r_{\star})$        &   -0.38 &   -0.10 &    0.09 &   -0.21 &    0.00 &    0.00 \\
$\alpha_D  ({\rm rad})$   &    1.65 &    1.41 &    1.89 &    1.30 &    1.02 &    1.45 \\
$\varphi_D  ({\rm rad})$  &    2.01 &    2.26 &    2.28 &    4.82 &    2.11 &    1.97 \\
 \cline{1-1}
$x_Q (r_{\star})$        &    0.05 &    0.32 &    0.19 &    0.37 &    0.60 &    0.00 \\
$y_Q (r_{\star})$        &    0.15 &    0.25 &   -0.18 &   -0.35 &    0.14 &    0.00 \\
$z_Q (r_{\star})$        &   -0.08 &   -0.24 &   -0.37 &   -0.27 &   -0.28 &   -0.12 \\
$\alpha_Q  ({\rm rad})$   &    2.21 &    2.30 &    0.59 &    1.48 &    0.88 &    0.79 \\
$\varphi_Q  ({\rm rad})$  &    1.92 &    2.46 &    5.97 &    4.83 &    5.75 &    4.91 \\
$B_{\rm Q}/B_{\rm D}$    &    5.97 &    5.42 &    3.15 &    2.12 &    9.82 &    6.86 \\
 \hline
$\chi_{\rm r}^2$       &    0.92 &    0.85 &    1.13 &    0.93 &    1.50 &    8.94 \\
\enddata
\tablecomments{Prefixes: R -- Riley, F -- force-free fields}
\end{deluxetable*}

The derivation of the FF field structure has a much higher
computational overhead (of order $\thicksim 10^2$ times more CPU
time per likelihood evaluation) than the analytic static vacuum
field models. Each FF field module runs in parallel using $12^3$
processors and therefore, the MCMC exploration is serially performed
(i.e., at every MCMC step only one point in the parameter space is
explored). The FF simulations {implement} a grid resolution
$0.04R_{\rm LC}$, which does not allow a simulation stellar surface
of radius smaller than $r_{\rm sim}=0.25R_{\rm LC}$. However, even
though this stellar radius is four to five times larger than the
actual one, it is still well inside the LC. Eventually, the
determination of the polar caps on the corresponding actual stellar
radius, $r_{\star}$, takes into account the field structure outside
$r_{\rm sim}=0.25R_{\rm LC}$ and the corresponding SVF structure
inside the stellar surface. The numerical FF solution (outside
$r_{\rm sim}$) is smoothly joined to the SVF (inside $r_{\rm sim}$)
through a linear ramp function.

Severe computational limitations do not allow a detailed exploration
of the parameter space for the FF regime, which would allow the
determination not only of the highest likelihood parameters but also
of the corresponding posterior distributions (e.g., corner plots).
Nonetheless, we ran in total 12 MCMC chains, 6 starting from the
highest likelihood parameter values of the $RV$ models and another 6
starting from different parameter values (i.e., not corresponding to
the highest-likelihood values) of the $RV$ models. In each of these
runs, we explored several hundreds to a few thousand points of the
parameter space while the accepted number of MCMC points were of the
order of a few hundred. The exploration in each of these MCMC chains
continued until either a likelihood value corresponding to
$\chi^2_{\rm r}\lesssim 0.8$ was found or a few thousand points in
the parameter space had been explored. Even though the accepted
number of points were inadequate to provide proper posterior
distributions, they were enough to provide likelihood values close
to the corresponding local maxima. Thus, in the FF cases, the MCMC
approach was mainly used as an optimization method.

In Table~\ref{tab:ff}, we present the highest likelihood model
parameters for six FF models. Each $RFn$ model is the highest
likelihood FF model among the models corresponding to the 2 MCMC
chains we ran for each $RVn$ model. Our results show that the first
five models ($RF1_{11}-RF5_{8}$) have adequately low $\chi_{\rm
r}^2$ values (i.e., of the order of 1) and acceptably describe the
data.

\begin{figure*}[!tbh]
\vspace{0.0in}
  \begin{center}
    \includegraphics[width=1.0\textwidth]{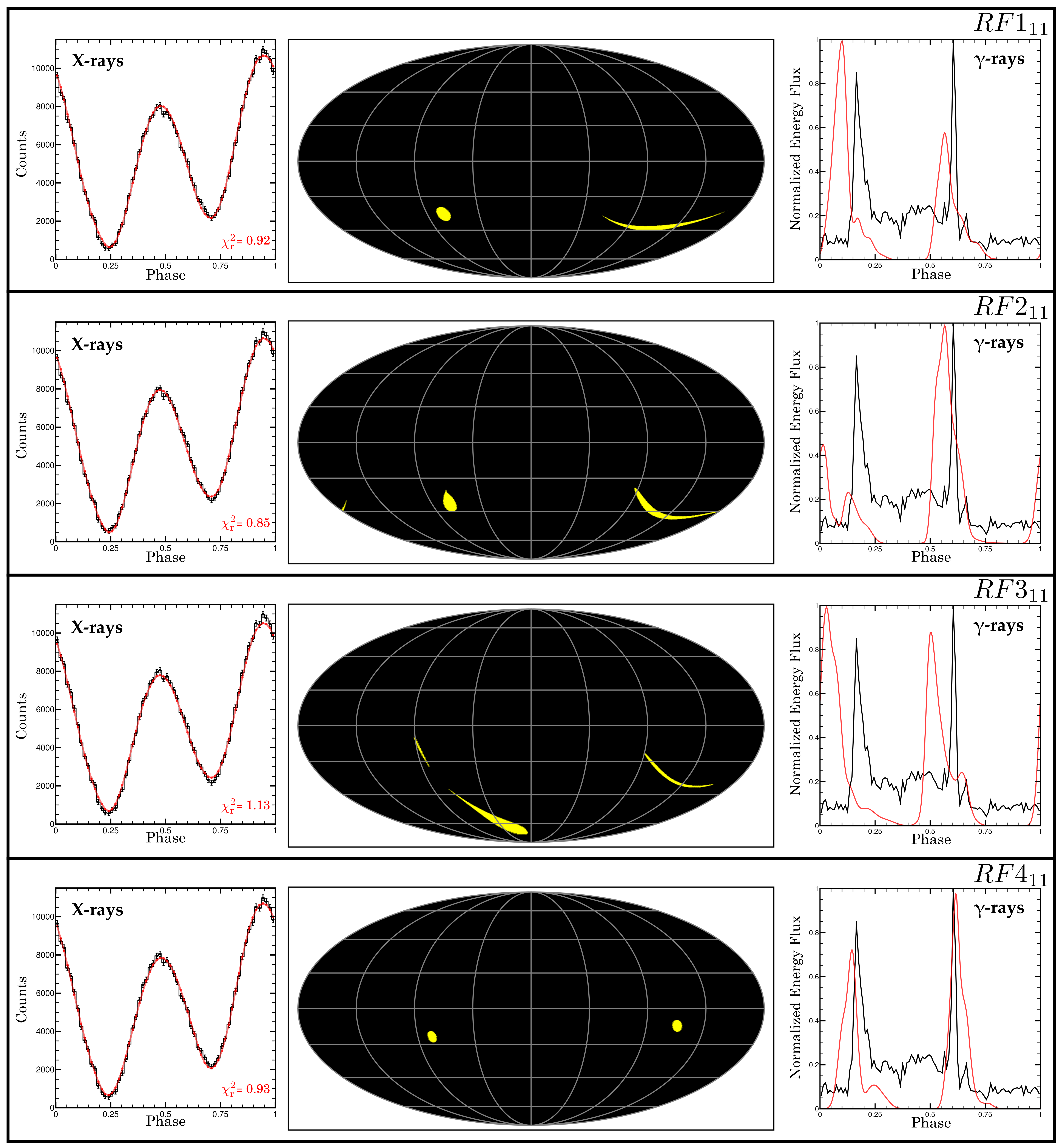}
  \end{center}
  \vspace{-0.25in}
  \caption{The $RF$ models with 11 free parameters. Each row corresponds
  to the indicated $RF$ model. The left-hand panels show the \nicer X-ray
  light curve (black lines with error-bars) together with the FF model
  X-ray light curve (red color). The corresponding $\chi^2_{\rm r}$ value
  are also denoted. The middle panels show the corresponding hot-spots
  (yellow regions) on the stellar surface. The right-hand panels show
  the \emph{Fermi-LAT} $\gamma$-ray light curve (black line) together with
  the model $\gamma$-ray light curve (red line).}
  \label{fig:06}
  \vspace{0.0in}
\end{figure*}

\begin{figure*}[!tbh]
\vspace{0.0in}
  \begin{center}
    \includegraphics[width=0.99\textwidth]{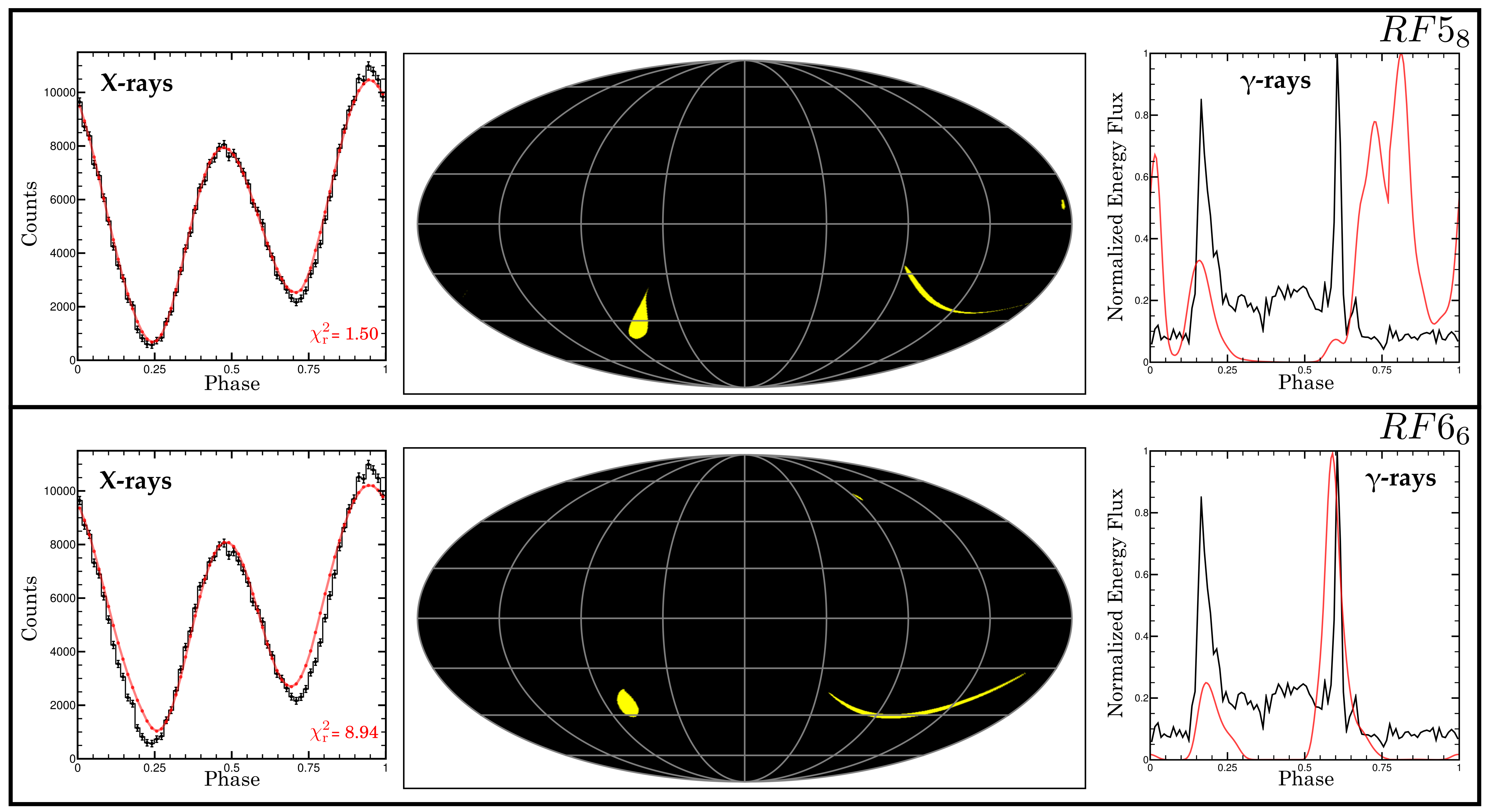}
  \end{center}
  \vspace{-0.2in}
  \caption{Similar to Fig.~\ref{fig:06} but for the $RF$ models with
  8 and 6 free parameters.}
  \label{fig:07}
  \vspace{0.0in}
\end{figure*}

In the left-hand column of Figs.~\ref{fig:06}--\ref{fig:07}, we plot
the FF X-ray light curves having the highest likelihood while in the
middle column, we plot the corresponding hot-spot regions on the
projected stellar surface. We see that the hot-spot patterns are
very similar to the corresponding $RV$ models, which implies that
the $RF$ models represent the same hot-spot families in the FF
regime. However, the areas of the FF hot-spot regions are, in
general, larger than those in SVF\footnote{{It is well known from
the pure central dipole solutions that the FF polar caps are larger
and shifted compared to those in the corresponding vacuum ones. This
is because the flowing currents not only increase the escaping
magnetic flux but the filed lines are also swept back due to
rotation.}}. Moreover, we note that a by-eye comparison indicates
that the $RF6_{6}$ model X-ray light curve is quite close to the
observed light curve and the corresponding high $\chi_{\rm r}^2$
value is due to the very small observational uncertainties.

For the $RF6$ model, we explored $\sim 4000$ points in the parameter
space getting $\sim 160$ accepted points in the MCMC chain. The
likelihood values stabilized to the ones quoted in
Table~\ref{tab:ff}. Nonetheless, as is an issue with all
higher-dimensional search problems, the limited number of accepted
points leaves unclear whether these optimized likelihoods
are a global maximum.

\subsubsection{Fermi-LAT $\gamma$-ray Light Curve}\label{FFfldfermi}

Our results indicate field degeneracies for both the SVF and FF
models. Even though the observed X-ray light-curve sets constraints
there are many different field structures that may adequately
describe the bolometric \emph{NICER} X-ray light curve. There are
not only different field parameters corresponding to different
hot-spot families (e.g., $RVn$ and $RFn$ families) of quite
different hot-spot patterns but also quite different field
parameters within the same families (see Fig.~\ref{fig:01}). Even
though there are models with higher and lower $\chi_{\rm r}^2$
values, the actual $\chi_{\rm r}^2$ values, for the majority of the
models, are of the order of unity and therefore, they are all
statistically acceptable, although some may be formally preferred,
e.g. by an F-test. Yet, as we show below, more pertinent is whether
the solutions can correctly describe the phasing of the $\gamma$-ray
light curves.

The variation of the magnetic field parameters affect the magnetic
field structure even near the LC. Usually at large distances, the
magnetic field geometry is dictated by the dipole moment parameters.
However, in MSPs like \mypsr the corresponding $R_{\rm LC}$ lies at
only $\approx 18 r_{\star}$, which implies that the magnetic moment
offsets, and the quadrupole strength would significantly affect the
field structure of the outer magnetosphere {i.e., near the LC}. The
offsets and the $f_{\rm Q}$ ratio would actually introduce
deviations of the field structure from central symmetry.

The field structure geometry in the outer magnetosphere and
especially near and beyond the LC affects the locus of the ECS and
therefore, the corresponding $\gamma$-ray emission. Assuming a pure
centered dipole field, the $\gamma$-ray emission observed by
\emph{Fermi-LAT} sets constraints on $\alpha$ and $\zeta$ (i.e., the
only parameters involved) although there are degeneracies.

The statistical quality of the X-ray signal is very high for all
phases in contrast to the $\gamma$-rays, and therefore, an accurate
reproduction of the X-ray light curve is required. On the other
hand, the phases of $\gamma$-ray peaks set constraints mainly on the
locus of the ECS, which is strongly related to the field structure
parameters. Other morphological characteristics of the $\gamma$-ray
light curves are less related to the field geometry. For instance,
the width of the pulses is mainly related to the width of the
dissipative magnetosphere region while the inter-peak and off-peak
emission is mainly related to the emission that is produced inside
the LC.

In \citet{2014ApJ...793...97K}, we introduced the so-called FIDO
models that reproduce the observed correlation between the
radio-lag, $\delta$ and peak-separation, $\Delta$ of young pulsars.
In \citet{2015ApJ...804...84B,2017ApJ...842...80K}, we expanded and
generalized the FIDO models revealing the dependence of the plasma
conductivity on the spin-down power $\dot{\mathcal{E}}$ that
reproduce not only the $\delta-\Delta$ correlation but also the
observed \emph{Fermi-LAT} spectra. In the FIDO model the $\gamma$-ray
emission that is produced near the ECS is due to curvature radiation
(CR).

\begin{figure}[!tbh]
\vspace{0.0in}
  \begin{center}
    \includegraphics[width=0.475\textwidth]{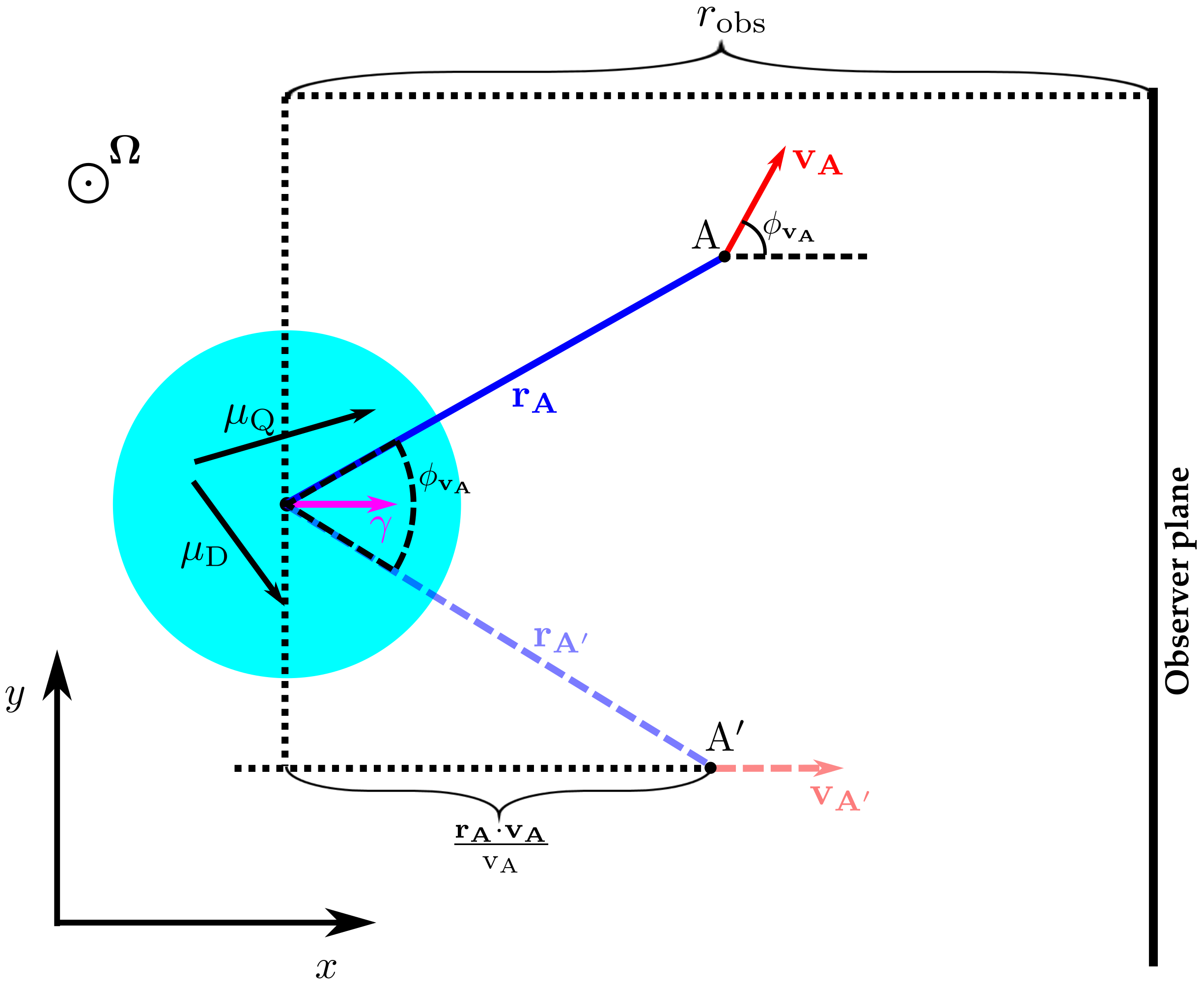}
  \end{center}
  \vspace{-0.2in}
  \caption{Schematic diagram, set in the equatorial plane for clarity,
  illustrating the self-consistent astrocentric time coordinates and
  synchronization of surface X-rays and magnetospheric $\gamma$-rays.}
  \label{fig:08}
  \vspace{0.0in}
\end{figure}

In this study, we adopt the approach we had followed in
\citep{2014ApJ...793...97K} for the derivation of the model
$\gamma$-ray light curves. We use the FF solutions adopting an
accelerating electric field component that reads
\begin{equation}
\label{eq:eparallel} E_{\parallel}=\frac{c (\boldsymbol{\nabla
\times B}) \cdot \boldsymbol{\hat{B}} }{4\pi\Sigma_{\rm c}}
\end{equation}
where $\boldsymbol{\hat{B}}$ is the local magnetic field unit vector
and $\Sigma_{\rm c}$ is the plasma conductivity. We have adopted
$\Sigma_{\rm c}=1\Omega$ constant everywhere; we note, however, that
the exact $\Sigma_{\rm c}$ value mainly affects the spectra and not
the $\gamma$-ray light curves as long as $\Sigma_{\rm c}$ is high
enough to produce a nearly-FF field solution.

We integrated test particle trajectories, using a $4^{\rm th}$ order
Runge-Kutta with constant step-size, that originate near the
separatrices of the open and closed field lines. More specifically,
for each $RF$ model, we calculated the polar-cap rims at
$r=0.5R_{\rm LC}$ and we integrated $10^6$ test particle
trajectories that originate at $r_{\rm tr}=0.5R_{\rm LC}$ within an
angle $3^{\circ}$ from the corresponding polar-cap rim. The particle
trajectories follow the Aristotelian electrodynamics\footnote{{In
Aristotelian electrodynamics the particle velocity is considered to
be always that of the locally defined asymptotic drift trajectory.}}
\citep{2012arXiv1205.3367G,2015AJ....149...33K,2015PhRvD..92b5029J}
and the particle energies are determined by both the energy gains
due to the encountered $E_{\parallel}$ and the energy losses due to
CR. A detailed description of the methodology is in
\citet{2014ApJ...793...97K} and \citet{2017ApJ...842...80K}. The
$E_{\parallel}$ and radiation energy loss rates govern particle
dynamics and energies. Photons are emitted along the direction of
test particle motion.

The construction of the model $\gamma$-ray sky-maps and light curves
requires a calculation of the photon phases, which should be
consistent with the phases of the X-ray photons. As mentioned above,
the $RV,~MV$, and $RF$ model parameters determine the orientation of
the stellar field at the times corresponding to phase zero of the
X-ray light curve. Following the formulation introduced in
\citep{2014ApJ...793...97K}, the phase of a $\gamma$-ray photon that
is emitted by a particle that lies at point A reads
\begin{equation}
\label{eq:grayphotonphase} \varphi_{\rm c}=\left(\Omega
t_{A}-\phi_{\mathbf{v_{A}}}-\frac{\mathbf{r_{A}}\cdot\mathbf{v_A}}{\rm
v_{A}}\frac{1}{R_{\rm
LC}}\right)\negthickspace\negthickspace\negthickspace\mod 2\pi
\end{equation}
where $t_A$ is the integration time corresponding to the point $A$
(assuming that the integration starts at $r_{\rm tr}$),
$\mathbf{v_A}, \mathbf{r_A}$ are the particle velocity and position
vectors at $A$, and $\phi_{\mathbf{v_A}}$ is the azimuth angle of
the velocity $\mathbf{v_A}$ with respect to the $x$ axis oriented
according to $\mathbf{\Omega}$ (see Fig.~\ref{fig:08}). We note that
the observer is located at $(r,\theta,\phi)=(r_{\rm obs},\zeta,0)$.

The radio ephemeris defines phase zero for both $\gamma$-ray and
X-ray observed light curves. We align our model X-ray light curve
with the observed X-ray light curve with phase zero as defined by
the radio phase zero \citep[as
in][]{2009ApJ...699.1171A,Bilous2019}. The phase of our model
$\gamma$-ray light curves are then computed from the orientation of
the field that produced the model X-ray light curve.

The phase $\varphi_{\rm c}$ is measured relative to a {fiducial}
photon that is emitted at the center of the star along the
line-of-sight direction. The emission of this fictitious photon
occurs at the time the observer receives the zero phase X-ray
photons in the image plane from {\tt{GIKS}}. Thus, the
synchronization of the X-ray and $\gamma$-ray photons requires a
modification of $\varphi_{\rm c}$ that takes into account the travel
time from the center of the star to the observer plane. This
significant nuance must be included to arrive at the correct
X-ray-to-$\gamma$-ray model phasing. We are not aware of this nuance
being accounted for properly in the literature. Importantly, this
accounts for the deviation in curved spacetime for the X-ray photon
trajectories in {\tt{GIKS}} reaching the image plane, and is
tantamount to an astrocentric clock correction. It is also important
to note that photons in any given observed phase may be emitted at
many mixed rotational phases owing to curved trajectories. Then, the
phases of the synchronized $\gamma$-ray photons read
\begin{equation}
\label{eq:synchronizedphase} \varphi_{\gamma}=\left(\varphi_{\rm
c}+\frac{\Omega\; r_{\rm obs}}{c}\right)
\negthickspace\negthickspace\negthickspace\mod 2\pi
\end{equation}
where the second term formulates the light travel time delay with
respect to the $\varphi_{\rm c}=0$ photons (see Fig.~\ref{fig:08}).

\begin{figure*}
\vspace{0.0in}
  \begin{center}
    \includegraphics[width=0.9\textwidth]{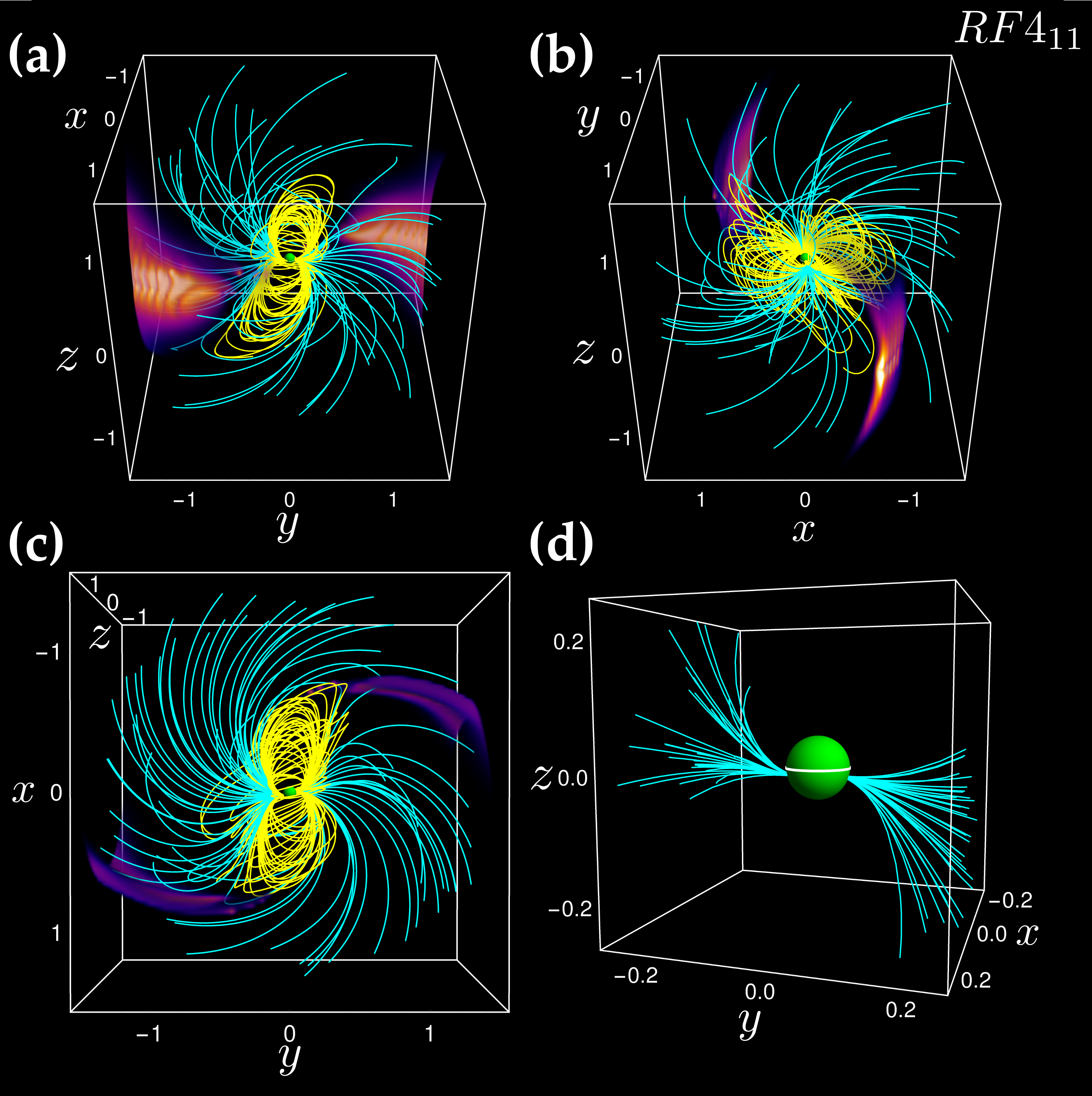}
  \end{center}
  \vspace{-0.1in}
  \caption{The field structure of the $RF4_{11}$ model. The open
  (closed) magnetic field lines are depicted by the cyan (yellow)
  colored lines. We note that the spin angular frequency vector
  lies along the $z$ axis and that the unit length is normalized 
  to $R_{\rm LC}\approx 18 r_{\star}$. In \textbf{(a)}-\textbf{(c)}, 
  the outer magnetosphere that reaches beyond the LC is shown from
  different points of view. The purple-to-white colored regions in
  \textbf{(a)}-\textbf{(c)} denote the high values of the
  ratio of the accelerating electric component over the total
  electric field that captures the reconnection region (i.e., ECS)
  that emerges from the tip of the closed region. In \textbf{(d)},
  the inner magnetosphere is shown up to $5r_{\star}$. Both
  polar caps (i.e., origins of the open magnetic field lines) clearly
  lie below the rotational equator (denoted by the white stripe).}
  \label{fig:new}
  \vspace{0.0in}
\end{figure*}

In the right-hand column of Figs.~\ref{fig:06}--\ref{fig:07}, we
plot the \emph{Fermi-LAT} $\gamma$-ray light curve (black lines)
together with the corresponding model $\gamma$-ray light curve (red
lines). Recall that the \emph{Fermi-LAT} $\gamma$-ray and \emph{NICER}
X-ray light curves are synchronized through radio ephemeris
\citep{2009ApJ...699.1171A,Bilous2019} while the model $\gamma$-ray
and X-ray light curves are synchronized through the phase
adjustments of Eq.~(\ref{eq:synchronizedphase}).

The model $\gamma$-ray light curves
(Figs.~\ref{fig:06}--\ref{fig:07}) indicate that a field
configuration that describes the observed X-ray light curve does not
necessary describe the observed $\gamma$-ray light curve. Thus,
combining X-ray and $\gamma$-ray light curves could reduce the
degeneracies in the possible field configurations.

The model $RF4_{11}$ does not only describe the X-ray light curve
but also describes very well the peaks of the $\gamma$-ray light
curve. The $\gamma$-ray light curves in models $RF1_{11}$ and
$RF3_{11}$ seem to have the correct peak separation but the peak
phases are quite off. On the other hand, the model $RF6_{6}$
captures well the peak phases of the $\gamma$-rays even though the
peak ratio P2/P1 is considerably larger than the observed one. The
disadvantage of this model is the rather poor statistics of the
X-ray fit.

\begin{figure}
\vspace{0.0in}
  \begin{center}
    \includegraphics[width=0.49\textwidth]{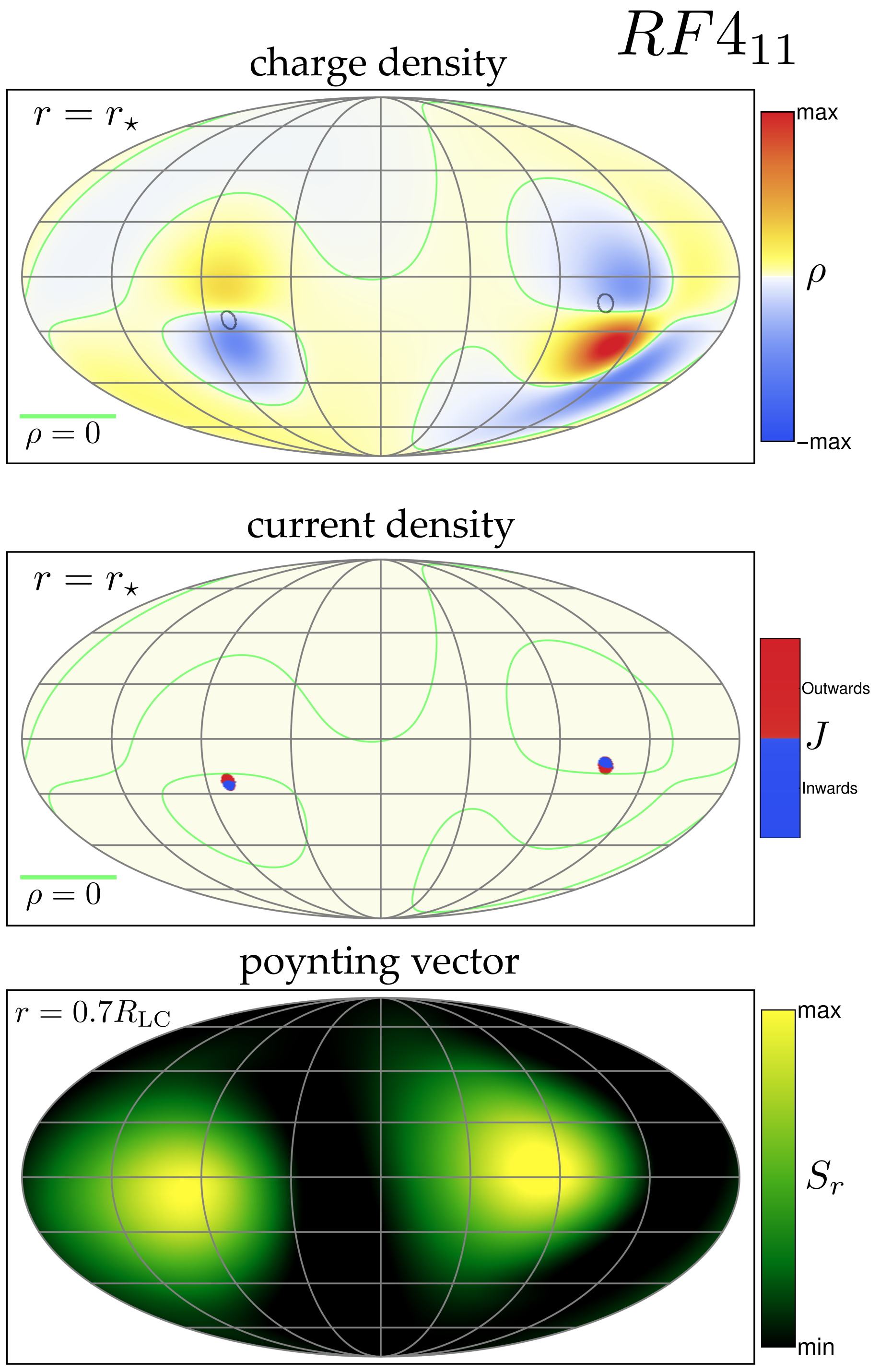}
  \end{center}
  \vspace{-0.2in}
  \caption{\textbf{Top panel:} The charge density on the stellar surface
  in the indicated color scale, for the $RF4_{11}$ model. The green line
  denotes the zero-charge line. \textbf{Middle panel:} The current density
  on the stellar surface. The current flows from/to the hot-spots
  (i.e., polar caps). The green line denotes the zero-charge line.
  \textbf{Bottom panel:} The strength of the radial component, $S_{\rm r}$,
  of the Poynting vector on a sphere at $r=0.7R_{\rm LC}$.}
  \label{fig:09}
  \vspace{0.0in}
\end{figure}

\begin{figure}
\vspace{0.0in}
  \begin{center}
    \includegraphics[width=0.49\textwidth]{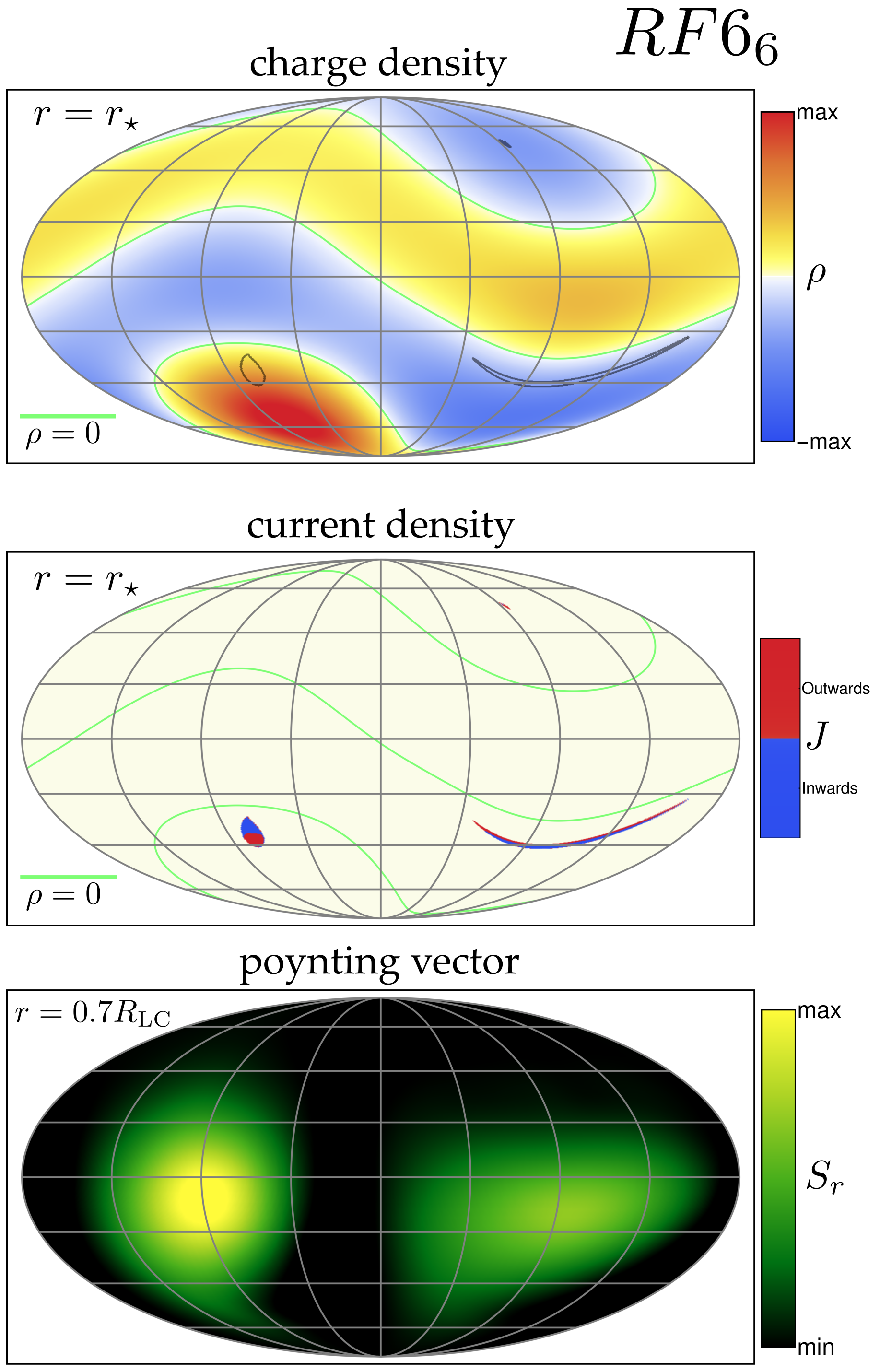}
  \end{center}
  \vspace{-0.2in}
  \caption{Similar to Fig.~\ref{fig:09}, but for the $RF6_{6}$ model.
  In this case, the two hot-spots lie in different charge polarities.
  The Poynting flux is asymmetric contributing to the rocket effect, which
  due to the rotation smoothing has mainly a component along the
  $\pmb{\Omega}$ axis.}
  \label{fig:10}
  \vspace{0.0in}
\end{figure}

{On the one hand, the generally low $\chi_{\rm r}^2$ values (i.e.,
$<1$) imply that the number of adopted parameters might be higher
than what is needed for the fitting of the bolometric X-ray light
curve. On the other hand, the high $\chi_{\rm r}^2$ value of the
$RF6_{6}$ model as well as the performance of the model $\gamma$-ray
light curves suggest that the eleven adopted parameters might be
necessary for the description of the fields that fit both the \nicer
X-ray and \emph{Fermi-LAT} $\gamma$-ray light curves.}

In Fig.~\ref{fig:new}, we present the field structure of the
$RF4_{11}$ model. The cyan colored lines indicate the open
magnetic field lines while the yellow colored lines the closed ones.
In panels (a)-(c), the outer magnetosphere (up to and beyond the LC)
is shown. In these panels, the high accelerating regions (i.e.,
colored 3D regions) where magnetic reconnection takes place (i.e.,
ECS region) and most of the $\gamma$-ray emission is produced are
also plotted. Panel (d) shows the inner magnetosphere up to
$5r_{\star}$. For clarity only the open magnetic field lines are
plotted. These lines originate from the corresponding two polar caps
(i.e, hot-spots), which both are located in the southern rotational
hemisphere, i.e., below the rotational equator which is indicated by
the white stripe.

\section{Conclusions}
\label{conclusions} In this study, we explored SVF and FF field
structures that fit the thermal X-ray light curve observed by
\emph{NICER} and \emph{Fermi-LAT} $\gamma$-ray light curve of \mypsR.

The adopted magnetic field model consists of the sum of an offset
dipole moment and an offset quadrupole $m=0$ moment and is fully
described by eleven parameters. {In our modeling, the stellar mass,
radius, and observer angle were fixed according to
\citetalias{Miller2019} and \citetalias{Riley2019}}. Moreover, the
same hot-spot temperature was adopted in agreement with the findings
of \citetalias{Miller2019} and \citetalias{Riley2019}.

Our results revealed a variety of families of SVF structures that
produce hot-spots (i.e., polar caps) that superbly describe (i.e.,
$\chi^2_{\rm r}\lesssim 1$) the bolometric X-ray light curve. The
rather fast calculations of the analytic SVF models allowed a
detailed MCMC exploration of the multimodal landscape corresponding
to the various hot-spot families. Our analysis indicated the
existence of configuration degeneracies not only between the
different families but also among the same families.

We also explored SVF models with additional constraints on the field
configuration that reduces the dimensionality of the parameter
space. Thus, we found models that have the dipole moment at the NS
center (i.e., no offset) and models where the quadrupole moment was
additionally restricted to lie along the $\pmb{\Omega}$ (i.e., $z$)
axis.

Starting from SVF solutions, we explored the parameter space of FF
field structures that describe the X-ray light curves. However, the
parameter space exploration in FF models was not as detailed as in
the SVF regime due to the extremely high computational demands.
Nonetheless, we found FF solutions that represent model families
similar to the SVF ones. For the vast majority of the FF models the
$\chi_{\rm r}^2$ values were of the order of unity.

Using the realistic FF field structures that describe the
X-ray light curve, we calculated synchronous $\gamma$-ray light
curves and compared them to those observed by \emph{Fermi-LAT}. For
the calculation of the $\gamma$-ray light curves, we implemented the
original FIDO models (\citealt{2014ApJ...793...97K}; see also
\citealt{2015ApJ...804...84B,2017ApJ...842...80K}) that set the
$\gamma$-ray emission near the ECS beyond the LC. The particle
emission is due to CR and therefore, it depends on the dissipative
electric fields the particles encounter and the geometric properties
(i.e., radius of curvature) of the particle trajectories.

An important nuance not reported elsewhere is the proper
synchronization of model surface X-rays and magnetospheric
$\gamma$-rays, {which is} necessary for a {correct} relative
description of the X-ray and $ \gamma$-ray light curves.

The comparison of the model $\gamma$-ray light curves with
the observed one {sets more constraints and potentially breaks the
field degeneracies}, revealing the field structure that
simultaneously describes both the X-ray and $\gamma$-ray light
curves. Our results identified one FF field structure that describes
the X-ray light curve well and provides an adequate model of the
main peaks of the corresponding $\gamma$-ray light curve. Our
results also identified another FF field structure with fewer free
parameters, which, even though it matches the $\gamma$-ray peak
phases well, has a rather poor statistical fit of the X-ray
light curve.

\section{Discussion}
\label{discussion}

The spin-down properties of J0030+0451 suggest a surface 
dipole field of $B_{\rm D} \sim 2-4 \times 10^8$~G. The radiation 
transport in the atmosphere starts being influenced by the magnetic 
field for $B_{\star}\gtrsim B_0=2\times 10^9 {\rm G}\approx\alpha_f^2 
B_{\rm cr}$, where $\alpha_f$ is the fine constant and 
$B_{\rm cr}\approx 4.4\times 10^{13}$ G is the critical field. 
Such influences impart anisotropy and depend on the local 
direction of the magnetic field.
Nonetheless, the 
magnetic atmosphere models start becoming essential for the 
description of anisotropies in radiative transport of thermal 
emission only for $B_{\star}\gtrsim B_1\approx2\times 10^{10}{\rm G}$
\citep[][W. Ho, private comm]{2001RvMP...73..629L,2014A&A...572A..69P}.
For the vast majority of models that do consistently 
sample regions of $B_{\star} \gtrsim B_{\rm D}$ across the hot spot(s) 
(see Figs.~\ref{fig:02}-\ref{fig:05}), the field does not exceed
$\sim 10 B_{\rm D}\sim B_0$, which is much
less than $B_1$. Therefore, magnetic influences on radiation
transport in these models are likely small in the \nicer band.
Our results indicate only one model (i.e.,
$RV5_{8}$) with $B_{\star\rm max}\approx 400B_{\rm D}$ 
that well exceeds the $B_1$ scale. 
However, even in the $RV5_{8}$ model the 
hot-spot area, where the $B$-field reaches high
values, is small and thus, likely does not substantially influence 
the X-ray light curve. Moreover, the actual
$B_{\rm D}$ value, assuming dipole plus quadrupole field components,
that is required to produce the observed spin-down power would be
smaller than the nominal $B_{\rm D}$ field mentioned above (i.e.,
$\sim 2-4 \times 10^8$~G$)$, which assumes only a pure centered
dipole \citep[see][]{2020MNRAS.tmp.3024P}. Especially for
the $RV5_{8}$ model, $B_{\rm Q}/B_{\rm D}=8.68$ and therefore, the quadrupole
component at the LC remains rather strong (i.e., only $\sim
0.5\approx 8.68/18$ times smaller than the corresponding dipole
one), which implies that the required $B_{\rm D}$ value would be
even smaller than the nominal pure centered dipole one. 
Thus, the non-magnetic atmosphere treatments in
\citetalias{Miller2019}, \citetalias{Riley2019} as well as in this 
study are likely secure. 
Finally, we note that the \nicer band that was used in 
\citet{Bogdanov2019} (i.e., $\epsilon: 0.25-1.45$ keV), 
\citetalias{Miller2019} (i.e., $\epsilon: 0.4-3$ keV), 
and \citetalias{Riley2019} (i.e., $\epsilon: 0.25-3$ keV)
would sample the electron cyclotron resonance for a magnetic field 
$B_{\rm cr}(\epsilon/m_{\rm e}c^2) \approx 2\times 10^{10}-2\times 10^{11}{\rm G}$,
where $m_{\rm e}$ the electron mass and $c$ the speed of light.
These values are exceeded only for the $RV5_{8}$ model. Nonetheless,
the corresponding $B_{\star}$ is quite non-uniform across the
polar caps and therefore, a cyclotron resonance spectral feature would be
smeared for this reason, in addition to thermal and Doppler broadening.

We note that the various SVF and FF models have been derived from
the maximization of the likelihood of the model X-ray light curve.
An F-test would easily compare the various models both those that
have the same number of parameters and those with different number
of parameters. However, it is doubtful how useful this information
would be with respect to the evaluation of the models on a practical
level. The vast majority of the models have $\chi_{\rm r}^2\approx
1$ and therefore, the determination of a model (i.e., field
structure) effectiveness should incorporate the behavior of
additional features (e.g., synchronous $\gamma$-ray light curve) that are
essential for the overall evaluation of the model potency. Focusing
solely on the statistical performance of the X-ray light curve might
be misleading. In the same manner, the $\chi_{\rm r}^2$
values corresponding to the SVF models (see Table~\ref{tab:vac})
appear, in general, lower than the FF ones (see Table~\ref{tab:ff}),
which implies a better performance of the SVF models compared to the
FF ones. However, even though the statistically accepted $\chi_{\rm
r}^2$ values of the FF models are higher than those of the SVF
models, the $\gamma$-ray light-curve description requires physically
realistic charges and currents of the FF models. There is no doubt
that the FF field structures are more physically accurate for
describing the field structure of \mypsR.

We note that in the current study we did not explore FF
field structures adopting as fixed the parameter values of
\citetalias{Miller2019}. The MCMC parameter space exploration of the
FF field structures is computationally very demanding and therefore,
we decided to focus on one parameter set (i.e.,
\citetalias{Riley2019}). This approach provided a clearer picture of
the underlying field degeneracies even under the same adopted fixed
parameter set, while it revealed the use of $\gamma$-rays is
essential in setting additional constraints. Further exploration of
the parameter space assuming fixed parameters that are taken from
either \citetalias{Miller2019} or \citetalias{Riley2019} is not
expected to alter the principal conclusions of this work.

A proper study of the magnetic field structure requires the
incorporation of all the parameters beyond the field-related ones.
Thus, an extended free parameter space would also include
$M_{\star}$, $r_{\star}$, $\zeta$, hot-spot temperatures, and source
distance. Increasing the number of free parameters could increase
the number of field structures that adequately fit the \emph{NICER}
X-ray light curve. However, the incorporation of the X-ray photon
energy dependencies may provide additional constraints. Our study
makes evident that the model evaluation should include the behavior
of the corresponding $\gamma$-ray light curve. Ultimately, the
adopted likelihood should appropriately combine measurements of the
performance of both the X-ray and the $\gamma$-ray light curves.
Even though this task is computationally cumbersome, it is feasible
in the near future with the tools we have developed. However, since
the statistical uncertainties on the X-ray light curve are much
smaller than those on the $\gamma$-ray light curve, the X-ray light
curve would dominate in such a joint fit. This problem has been
studied in joint fits of $\gamma$-ray and radio pulsar light curves,
and requires careful treatment \citep[][Seyffert et al. 2020, in
prep.]{2014ApJS..213....6J}.

In our study, the hot-spot regions in FF magnetospheres are
considered to coincide with the entire polar-cap regions. However,
the polar-cap regions in FF field structures support a variety of
current density distributions. The integrated total current is zero
and therefore, a part of the polar cap is the origin of the direct
current while the rest of the polar cap is the origin of the return
current. The current patterns in the pure central dipole depend
solely on the magnetic obliquity, $\alpha$. {The current and charge
patterns across the polar caps are critical elements needed to
ultimately determine the surface heating and temperature
distribution. However, the polar-cap heating patterns require
detailed modeling of the particles that return to the surface from
the pair cascades, which is a small subset of the outflowing
particles, and the heating of the atmosphere by the returning
particles \citep{2019ApJ...872..162B,2020A&A...641A..15S}.
Therefore, the polar-cap temperature distributions cannot be
estimated directly from the current and charge patterns, other than
to note that regions of space-like ($J/\rho c > 1$) and return
($J/\rho c < 0$) currents enable pair cascades that will heat the
polar caps but regions of time-like current ($0 < J/\rho c < 1$)
{will not have pair-cascades
\citep{2013MNRAS.429...20T}\footnote{{Even though pair cascades are
not activated near the stellar surface above the time-like current
regions, it is still unknown whether pairs that originate from the
outer magnetosphere and especially from the regions where the
time-like current flows cross the zero charge surface return to the
polar cap.}}.}} {This implies that a more accurate treatment should
consider different temperatures for the regions of these different
current components within the same polar cap.}

{In the top and middle panels of Fig.~\ref{fig:09}, we plot, in the
indicated color scales, the charge and current patterns,
respectively on the projected stellar surface for the $RF4_{11}$
model. In this case, both polar caps are located within the
negatively charged magnetosphere regions.} However, in these field
configurations, the zero charge line can pass through the polar caps
or the polar caps can lie in different charge polarities.
Figure~\ref{fig:10} shows similarly (to Fig.~\ref{fig:09}) the
charge and current density patterns for the $RF6_6$ models. In this
case, the two hot-spots (i.e., polar caps) are located in regions of
opposite charge polarities and therefore, in one polar cap the
return current (i.e., $J/\rho c<0$) corresponds to the inward
current while in the other polar cap to the outward current.

The determination of the magnetic field structure through the
thermal X-ray and $\gamma$-ray light-curve constraints and the
possible field degeneracies would provide not only a deeper
understanding of the corresponding emission processes but also
confidence regarding the validity of the current emission modeling.

{However, the implications of this study are much broader. Our
modeling informs} on the internal field of the MSP, albeit as an
external boundary condition. How might the multipolar field
configurations arise? One possibility is through the migration of
the neutron and proton superfluid vortices as detailed in
\citet{1991ApJ...366..261R,1991ApJ...382..576R,1993ApJ...408..179C}.
Interestingly, such a migration should produce large values for
$z_{\rm D}$, and rather small values for $x_{\rm D},~y_{\rm D}$, in
contrast to our model solutions. Such field migration also requires
a small critical strain of the crust, in contrast to large one
suggested by recent simulations \citep{2009PhRvL.102s1102H}. Another
possibility is burial of the field
\citep[e.g.,][]{2004MNRAS.351..569P} during the recycling process,
but it is not clear that a burial mechanism is effective for
permanent field reduction in MSPs or that multipolar external fields
would necessarily result. Finally, we note that other
unknown internal field processes, analogous to those advanced for
young pulsars
\citep[e.g.,][]{2014MNRAS.444.3198G,2018ApJ...852...21G}, but
operating on much longer timescales might result in such multipolar
fields.

The field structure on the stellar surface may considerably affect
the efficiency of the pair production that takes place above the
polar caps near the stellar surface \citep{2015ApJ...810..144T}. For
instance, higher pair multiplicity could have important implications
for the locally measured energetic positrons excess
\citep{2015ApJ...807..130V} or the efficiency of the pulsed
$\gamma$-ray emission \citep{2018ApJ...857...44K}.

The determination of the absolute phase offset of the model X-ray
and $\gamma$-ray light curves in a particular magnetic field
solution will also produce strong constraints on the altitude of the
radio emission, since we know that the thermal X-rays are radiated
at the NS surface. Radio emission heights have traditionally been
estimated assuming a centered dipole field and some geometric model
for the radio beam \citep{2004Dyks}. Our determination of the field
structure near the NS together with the absolute phase positions of
both X-ray, $\gamma$-ray and radio peaks will provide the altitude
of the radio emission for a given radio beam geometry, thus
constraining radio emission modeling.

Moreover, the complicated field structure would provide insight into
the mechanisms for field evolution in MSPs \citep{2013Vigano}.
Finally, the asymmetric fields that seem to be present at least in
MSPs would contribute to the rocket effect
\citep{1975Harrison,2020MNRAS.tmp.3024P}. {Such acceleration could
be an unmodeled systematic in using MSPs to map the galactic
potential.} In the bottom panels of
Figs.~\ref{fig:09}--\ref{fig:10}, we plot the Poynting flux, in the
indicated color scale, for the $RF4_{11}$ and $RF6_6$ models,
respectively on the Mollweide projection of the sphere $r=0.7R_{\rm
LC}$. The asymmetric radiation pattern (more pronounced in $RF6_6$)
will accelerate the star towards the opposite direction. However, we
note that only the asymmetric component along $\pmb{\Omega}$ will
contribute to the net-linear acceleration of the star since all the
other components will be, on average, canceled out due to the fast
rotation.
\vspace{0.1in}

\acknowledgements {We thank an anonymous referee for the
constructive comments that helped us improving the clarity of the
paper. We acknowledge helpful discussions with Wynn Ho, Slavko
Bogdanov, and Zaven Arzoumanian. {We especially thank Anna Watts,
Thomas Riley, Anna Bilous, Cole Miller, and Fred Lamb for carefully
reading and providing helpful feedback on an earlier version of this
manuscript. We would also like to thank the International Space
Science Institute (ISSI) for providing financial support and for the
organization of the meeting of the ISSI Team that was led by I.
Contopoulos and D. Kazanas.} Z.W. is supported by the NASA
postdoctoral program. A.K.H. and D.K are supported by the {\it
Fermi} Guest Investigator program. This work has made use of the
NASA Astrophysics Data System.}


\begin{thebibliography}{}
\expandafter\ifx\csname natexlab\endcsname\relax\def\natexlab#1{#1}\fi
\providecommand{\url}[1]{\href{#1}{#1}}
\providecommand{\dodoi}[1]{doi:~\href{http://doi.org/#1}{\nolinkurl{#1}}}
\providecommand{\doeprint}[1]{\href{http://ascl.net/#1}{\nolinkurl{http://ascl.net/#1}}}
\providecommand{\doarXiv}[1]{\href{https://arxiv.org/abs/#1}{\nolinkurl{https://arxiv.org/abs/#1}}}

\bibitem[{{Abdo} {et~al.}(2009{\natexlab{a}}){Abdo}, {Ackermann}, {Atwood},
  {Axelsson}, {Baldini}, {Ballet}, {Barbiellini}, {Bastieri}, {Battelino},
  {Baughman}, {Bechtol}, {Bellazzini}, {Berenji}, {Bloom}, {Bonamente},
  {Borgland }, {Bregeon}, {Brez}, {Brigida}, {Bruel}, {Burnett}, {Caliandro},
  {Cameron}, {Caraveo}, {Casandjian}, {Cecchi}, {Charles}, {Chekhtman},
  {Cheung}, {Chiang}, {Ciprini}, {Claus}, {Cognard}, {Cohen-Tanugi},
  {Cominsky}, {Conrad}, {Cutini}, {Dermer}, {de Angelis}, {de Palma}, {Digel},
  {Dormody}, {Silva}, {Drell}, {Dubois}, {Dumora}, {Farnier}, {Favuzzi},
  {Focke}, {Frailis}, {Fukazawa}, {Funk}, {Fusco}, {Gargano}, {Gasparrini},
  {Gehrels}, {Germani}, {Giebels}, {Giglietto}, {Giordano}, {Glanzman},
  {Godfrey}, {Grenier}, {Grondin}, {Grove}, {Guillemot}, {Guiriec}, {Hanabata},
  {Harding}, {Hayashida}, {Hays}, {Hughes}, {J{\'o}hannesson}, {Johnson},
  {Johnson}, {Johnson}, {Johnson}, {Kamae}, {Katagiri}, {Kataoka}, {Kawai},
  {Kerr}, {Kn{\"o}dlseder}, {Kocian}, {Komin}, {Kuehn}, {Kuss}, {Lande},
  {Latronico}, {Lee}, {Lemoine-Goumard}, {Longo}, {Loparco}, {Lott},
  {Lovellette}, {Lubrano}, {Madejski}, {Makeev}, {Marelli}, {Mazziotta},
  {McConville}, {McEnery}, {Meurer}, {Michelson}, {Mitthumsiri}, {Mizuno},
  {Moiseev}, {Monte}, {Monzani}, {Morselli}, {Moskalenko}, {Murgia}, {Nolan},
  {Nuss}, {Ohsugi}, {Omodei}, {Orlando}, {Ormes}, {Pancrazi}, {Paneque},
  {Panetta}, {Parent}, {Pepe}, {Pesce-Rollins}, {Piron}, {Porter}, {Rain{\`o}},
  {Rando}, {Razzano}, {Reimer}, {Reimer}, {Reposeur}, {Ritz}, {Rochester},
  {Rodriguez}, {Romani}, {Ryde}, {Sadrozinski}, {Sanchez}, {Sander},
  {Parkinson}, {Sgr{\`o}}, {Siskind}, {Smith}, {Smith}, {Spandre}, {Spinelli},
  {Starck}, {Strickman}, {Suson}, {Tajima}, {Takahashi}, {Tanaka}, {Thayer},
  {Thayer}, {Theureau}, {Thompson}, {Tibaldo}, {Torres}, {Tosti}, {Tramacere},
  {Uchiyama}, {Usher}, {Van Etten}, {Vilchez}, {Vitale}, {Waite}, {Watters},
  {Webb}, {Wood}, {Ylinen}, \& {Ziegler}}]{Abdo2009}
{Abdo}, A.~A., {Ackermann}, M., {Atwood}, W.~B., {et~al.} 2009{\natexlab{a}},
  \apj, 699, 1171, \dodoi{10.1088/0004-637X/699/2/1171}

\bibitem[{{Abdo} {et~al.}(2009{\natexlab{b}}){Abdo}, {Ackermann}, {Atwood},
  {Axelsson}, {Baldini}, {Ballet}, {Barbiellini}, {Bastieri}, {Battelino},
  {Baughman}, {Bechtol}, {Bellazzini}, {Berenji}, {Bloom}, {Bonamente},
  {Borgland }, {Bregeon}, {Brez}, {Brigida}, {Bruel}, {Burnett}, {Caliandro},
  {Cameron}, {Caraveo}, {Casandjian}, {Cecchi}, {Charles}, {Chekhtman},
  {Cheung}, {Chiang}, {Ciprini}, {Claus}, {Cognard}, {Cohen-Tanugi},
  {Cominsky}, {Conrad}, {Cutini}, {Dermer}, {de Angelis}, {de Palma}, {Digel},
  {Dormody}, {Silva}, {Drell}, {Dubois}, {Dumora}, {Farnier}, {Favuzzi},
  {Focke}, {Frailis}, {Fukazawa}, {Funk}, {Fusco}, {Gargano}, {Gasparrini},
  {Gehrels}, {Germani}, {Giebels}, {Giglietto}, {Giordano}, {Glanzman},
  {Godfrey}, {Grenier}, {Grondin}, {Grove}, {Guillemot}, {Guiriec}, {Hanabata},
  {Harding}, {Hayashida}, {Hays}, {Hughes}, {J{\'o}hannesson}, {Johnson},
  {Johnson}, {Johnson}, {Johnson}, {Kamae}, {Katagiri}, {Kataoka}, {Kawai},
  {Kerr}, {Kn{\"o}dlseder}, {Kocian}, {Komin}, {Kuehn}, {Kuss}, {Lande},
  {Latronico}, {Lee}, {Lemoine-Goumard}, {Longo}, {Loparco}, {Lott},
  {Lovellette}, {Lubrano}, {Madejski}, {Makeev}, {Marelli}, {Mazziotta},
  {McConville}, {McEnery}, {Meurer}, {Michelson}, {Mitthumsiri}, {Mizuno},
  {Moiseev}, {Monte}, {Monzani}, {Morselli}, {Moskalenko}, {Murgia}, {Nolan},
  {Nuss}, {Ohsugi}, {Omodei}, {Orlando}, {Ormes}, {Pancrazi}, {Paneque},
  {Panetta}, {Parent}, {Pepe}, {Pesce-Rollins}, {Piron}, {Porter}, {Rain{\`o}},
  {Rando}, {Razzano}, {Reimer}, {Reimer}, {Reposeur}, {Ritz}, {Rochester},
  {Rodriguez}, {Romani}, {Ryde}, {Sadrozinski}, {Sanchez}, {Sander},
  {Parkinson}, {Sgr{\`o}}, {Siskind}, {Smith}, {Smith}, {Spandre}, {Spinelli},
  {Starck}, {Strickman}, {Suson}, {Tajima}, {Takahashi}, {Tanaka}, {Thayer},
  {Thayer}, {Theureau}, {Thompson}, {Tibaldo}, {Torres}, {Tosti}, {Tramacere},
  {Uchiyama}, {Usher}, {Van Etten}, {Vilchez}, {Vitale}, {Waite}, {Watters},
  {Webb}, {Wood}, {Ylinen}, \& {Ziegler}}]{2009ApJ...699.1171A}
---. 2009{\natexlab{b}}, \apj, 699, 1171, \dodoi{10.1088/0004-637X/699/2/1171}

\bibitem[{{Bai} \& {Spitkovsky}(2010)}]{2010ApJ...715.1282B}
{Bai}, X.-N., \& {Spitkovsky}, A. 2010, \apj, 715, 1282,
  \dodoi{10.1088/0004-637X/715/2/1282}

\bibitem[{{Baub{\"o}ck} {et~al.}(2013){Baub{\"o}ck}, {Berti}, {Psaltis}, \&
  {{\"O}zel}}]{2013ApJ...777...68B}
{Baub{\"o}ck}, M., {Berti}, E., {Psaltis}, D., \& {{\"O}zel}, F. 2013, \apj,
  777, 68, \dodoi{10.1088/0004-637X/777/1/68}

\bibitem[{{Baub{\"o}ck} {et~al.}(2019){Baub{\"o}ck}, {Psaltis}, \&
  {{\"O}zel}}]{2019ApJ...872..162B}
{Baub{\"o}ck}, M., {Psaltis}, D., \& {{\"O}zel}, F. 2019, \apj, 872, 162,
  \dodoi{10.3847/1538-4357/aafe08}

\bibitem[{{Baub{\"o}ck} {et~al.}(2012){Baub{\"o}ck}, {Psaltis}, {{\"O}zel}, \&
  {Johannsen}}]{2012ApJ...753..175B}
{Baub{\"o}ck}, M., {Psaltis}, D., {{\"O}zel}, F., \& {Johannsen}, T. 2012,
  \apj, 753, 175, \dodoi{10.1088/0004-637X/753/2/175}

\bibitem[{{Bilous} {et~al.}(2019){Bilous}, {Watts}, {Harding}, {Riley},
  {Arzoumanian}, {Bogdanov}, {Gendreau}, {Ray}, {Guillot}, {Ho}, \&
  {Chakrabarty}}]{Bilous2019}
{Bilous}, A.~V., {Watts}, A.~L., {Harding}, A.~K., {et~al.} 2019, \apjl, 887,
  L23, \dodoi{10.3847/2041-8213/ab53e7}

\bibitem[{{Bogdanov} {et~al.}(2019{\natexlab{a}}){Bogdanov}, {Lamb},
  {Mahmoodifar}, {Miller}, {Morsink}, {Riley}, {Strohmayer}, {Tung}, {Watts},
  {Dittmann}, {Chakrabarty}, {Guillot}, {Arzoumanian}, \&
  {Gendreau}}]{2019ApJ...887L..26B}
{Bogdanov}, S., {Lamb}, F.~K., {Mahmoodifar}, S., {et~al.} 2019{\natexlab{a}},
  \apjl, 887, L26, \dodoi{10.3847/2041-8213/ab5968}

\bibitem[{{Bogdanov} {et~al.}(2019{\natexlab{b}}){Bogdanov}, {Guillot}, {Ray},
  {Wolff}, {Chakrabarty}, {Ho}, {Kerr}, {Lamb}, {Lommen}, {Ludlam}, {Milburn},
  {Montano}, {Miller}, {Baub{\"o}ck}, {{\"O}zel}, {Psaltis}, {Remillard},
  {Riley}, {Steiner}, {Strohmayer}, {Watts}, {Wood}, {Zeldes}, {Enoto},
  {Okajima}, {Kellogg}, {Baker}, {Markwardt}, {Arzoumanian}, \&
  {Gendreau}}]{Bogdanov2019}
{Bogdanov}, S., {Guillot}, S., {Ray}, P.~S., {et~al.} 2019{\natexlab{b}},
  \apjl, 887, L25, \dodoi{10.3847/2041-8213/ab53eb}

\bibitem[{{Brambilla} {et~al.}(2015){Brambilla}, {Kalapotharakos}, {Harding},
  \& {Kazanas}}]{2015ApJ...804...84B}
{Brambilla}, G., {Kalapotharakos}, C., {Harding}, A.~K., \& {Kazanas}, D. 2015,
  \apj, 804, 84, \dodoi{10.1088/0004-637X/804/2/84}

\bibitem[{{Cerutti} {et~al.}(2016){Cerutti}, {Philippov}, \&
  {Spitkovsky}}]{2016MNRAS.457.2401C}
{Cerutti}, B., {Philippov}, A.~A., \& {Spitkovsky}, A. 2016, \mnras, 457, 2401,
  \dodoi{10.1093/mnras/stw124}

\bibitem[{{Chen} {et~al.}(2020){Chen}, {Yuan}, \& {Vasilopoulos}}]{Chen2020}
{Chen}, A.~Y., {Yuan}, Y., \& {Vasilopoulos}, G. 2020, \apjl, 893, L38,
  \dodoi{10.3847/2041-8213/ab85c5}

\bibitem[{{Chen} \& {Ruderman}(1993)}]{1993ApJ...408..179C}
{Chen}, K., \& {Ruderman}, M. 1993, \apj, 408, 179, \dodoi{10.1086/172578}

\bibitem[{{Contopoulos} \& {Kalapotharakos}(2010)}]{2010MNRAS.404..767C}
{Contopoulos}, I., \& {Kalapotharakos}, C. 2010, \mnras, 404, 767,
  \dodoi{10.1111/j.1365-2966.2010.16338.x}

\bibitem[{{Contopoulos} {et~al.}(1999){Contopoulos}, {Kazanas}, \&
  {Fendt}}]{1999ApJ...511..351C}
{Contopoulos}, I., {Kazanas}, D., \& {Fendt}, C. 1999, \apj, 511, 351,
  \dodoi{10.1086/306652}

\bibitem[{{Dyks} {et~al.}(2004){Dyks}, {Rudak}, \& {Harding}}]{2004Dyks}
{Dyks}, J., {Rudak}, B., \& {Harding}, A.~K. 2004, \apj, 607, 939,
  \dodoi{10.1086/383587}

\bibitem[{{Gendreau} {et~al.}(2016){Gendreau}, {Arzoumanian}, {Adkins},
  {Albert}, {Anders}, {Aylward}, {Baker}, {Balsamo}, {Bamford}, {Benegalrao},
  {Berry}, {Bhalwani}, {Black}, {Blaurock}, {Bronke}, {Brown}, {Budinoff},
  {Cantwell}, {Cazeau}, {Chen}, {Clement}, {Colangelo}, {Coleman},
  {Coopersmith}, {Dehaven}, {Doty}, {Egan}, {Enoto}, {Fan}, {Ferro}, {Foster},
  {Galassi}, {Gallo}, {Green}, {Grosh}, {Ha}, {Hasouneh}, {Heefner}, {Hestnes},
  {Hoge}, {Jacobs}, {J{\o}rgensen}, {Kaiser}, {Kellogg}, {Kenyon}, {Koenecke},
  {Kozon}, {LaMarr}, {Lambertson}, {Larson}, {Lentine}, {Lewis}, {Lilly},
  {Liu}, {Malonis}, {Manthripragada}, {Markwardt}, {Matonak}, {Mcginnis},
  {Miller}, {Mitchell}, {Mitchell}, {Mohammed}, {Monroe}, {Montt de Garcia},
  {Mul{\'e}}, {Nagao}, {Ngo}, {Norris}, {Norwood}, {Novotka}, {Okajima},
  {Olsen}, {Onyeachu}, {Orosco}, {Peterson}, {Pevear}, {Pham}, {Pollard},
  {Pope}, {Powers}, {Powers}, {Price}, {Prigozhin}, {Ramirez}, {Reid},
  {Remillard}, {Rogstad}, {Rosecrans}, {Rowe}, {Sager}, {Sanders}, {Savadkin},
  {Saylor}, {Schaeffer}, {Schweiss}, {Semper}, {Serlemitsos}, {Shackelford},
  {Soong}, {Struebel}, {Vezie}, {Villasenor}, {Winternitz}, {Wofford},
  {Wright}, {Yang}, \& {Yu}}]{2016SPIE.9905E..1HG}
{Gendreau}, K.~C., {Arzoumanian}, Z., {Adkins}, P.~W., {et~al.} 2016, Society
  of Photo-Optical Instrumentation Engineers (SPIE) Conference Series, Vol.
  9905, {The Neutron star Interior Composition Explorer (NICER): design and
  development}, 99051H, \dodoi{10.1117/12.2231304}

\bibitem[{{Geppert} \& {Vigan{\`o}}(2014)}]{2014MNRAS.444.3198G}
{Geppert}, U., \& {Vigan{\`o}}, D. 2014, \mnras, 444, 3198,
  \dodoi{10.1093/mnras/stu1675}

\bibitem[{{Gonthier} \& {Harding}(1994)}]{1994ApJ...425..767G}
{Gonthier}, P.~L., \& {Harding}, A.~K. 1994, \apj, 425, 767,
  \dodoi{10.1086/174020}

\bibitem[{{Goodman} \& {Weare}(2010)}]{2010CAMCS...5...65G}
{Goodman}, J., \& {Weare}, J. 2010, Communications in Applied Mathematics and
  Computational Science, 5, 65, \dodoi{10.2140/camcos.2010.5.65}

\bibitem[{{Gourgouliatos} \& {Hollerbach}(2018)}]{2018ApJ...852...21G}
{Gourgouliatos}, K.~N., \& {Hollerbach}, R. 2018, \apj, 852, 21,
  \dodoi{10.3847/1538-4357/aa9d93}

\bibitem[{{Gralla} {et~al.}(2017){Gralla}, {Lupsasca}, \&
  {Philippov}}]{2017ApJ...851..137G}
{Gralla}, S.~E., {Lupsasca}, A., \& {Philippov}, A. 2017, \apj, 851, 137,
  \dodoi{10.3847/1538-4357/aa978d}

\bibitem[{{Gruzinov}(2012)}]{2012arXiv1205.3367G}
{Gruzinov}, A. 2012, arXiv e-prints.
\newblock \doarXiv{1205.3367}

\bibitem[{{Harding} \& {Muslimov}(2001)}]{Harding2001}
{Harding}, A.~K., \& {Muslimov}, A.~G. 2001, \apj, 556, 987,
  \dodoi{10.1086/321589}

\bibitem[{{Harrison} \& {Tademaru}(1975)}]{1975Harrison}
{Harrison}, E.~R., \& {Tademaru}, E. 1975, \apj, 201, 447,
  \dodoi{10.1086/153907}

\bibitem[{{Horowitz} \& {Kadau}(2009)}]{2009PhRvL.102s1102H}
{Horowitz}, C.~J., \& {Kadau}, K. 2009, \prl, 102, 191102,
  \dodoi{10.1103/PhysRevLett.102.191102}

\bibitem[{{Jacobson}(2015)}]{2015PhRvD..92b5029J}
{Jacobson}, T. 2015, \prd, 92, 025029, \dodoi{10.1103/PhysRevD.92.025029}

\bibitem[{{Johannsen} \& {Psaltis}(2010)}]{2010ApJ...718..446J}
{Johannsen}, T., \& {Psaltis}, D. 2010, \apj, 718, 446,
  \dodoi{10.1088/0004-637X/718/1/446}

\bibitem[{{Johnson} {et~al.}(2014){Johnson}, {Venter}, {Harding}, {Guillemot},
  {Smith}, {Kramer}, {{\c C}elik}, {den Hartog}, {Ferrara}, {Hou}, {Lande}, \&
  {Ray}}]{2014ApJS..213....6J}
{Johnson}, T.~J., {Venter}, C., {Harding}, A.~K., {et~al.} 2014, \apjs, 213, 6,
  \dodoi{10.1088/0067-0049/213/1/6}

\bibitem[{{Kalapotharakos} {et~al.}(2018){Kalapotharakos}, {Brambilla},
  {Timokhin}, {Harding}, \& {Kazanas}}]{2018ApJ...857...44K}
{Kalapotharakos}, C., {Brambilla}, G., {Timokhin}, A., {Harding}, A.~K., \&
  {Kazanas}, D. 2018, \apj, 857, 44, \dodoi{10.3847/1538-4357/aab550}

\bibitem[{{Kalapotharakos} \& {Contopoulos}(2009)}]{2009A&A...496..495K}
{Kalapotharakos}, C., \& {Contopoulos}, I. 2009, \aap, 496, 495,
  \dodoi{10.1051/0004-6361:200810281}

\bibitem[{{Kalapotharakos} {et~al.}(2014){Kalapotharakos}, {Harding}, \&
  {Kazanas}}]{2014ApJ...793...97K}
{Kalapotharakos}, C., {Harding}, A.~K., \& {Kazanas}, D. 2014, \apj, 793, 97,
  \dodoi{10.1088/0004-637X/793/2/97}

\bibitem[{{Kalapotharakos} {et~al.}(2017){Kalapotharakos}, {Harding},
  {Kazanas}, \& {Brambilla}}]{2017ApJ...842...80K}
{Kalapotharakos}, C., {Harding}, A.~K., {Kazanas}, D., \& {Brambilla}, G. 2017,
  \apj, 842, 80, \dodoi{10.3847/1538-4357/aa713a}

\bibitem[{{Kalapotharakos} {et~al.}(2019){Kalapotharakos}, {Harding},
  {Kazanas}, \& {Wadiasingh}}]{2019ApJ...883L...4K}
{Kalapotharakos}, C., {Harding}, A.~K., {Kazanas}, D., \& {Wadiasingh}, Z.
  2019, \apjl, 883, L4, \dodoi{10.3847/2041-8213/ab3e0a}

\bibitem[{{Kalapotharakos} {et~al.}(2012){Kalapotharakos}, {Kazanas},
  {Harding}, \& {Contopoulos}}]{Kala2012}
{Kalapotharakos}, C., {Kazanas}, D., {Harding}, A., \& {Contopoulos}, I. 2012,
  \apj, 749, 2, \dodoi{10.1088/0004-637X/749/1/2}

\bibitem[{{Kelner} {et~al.}(2015){Kelner}, {Prosekin}, \&
  {Aharonian}}]{2015AJ....149...33K}
{Kelner}, S.~R., {Prosekin}, A.~Y., \& {Aharonian}, F.~A. 2015, \aj, 149, 33,
  \dodoi{10.1088/0004-6256/149/1/33}

\bibitem[{{Lai}(2001)}]{2001RvMP...73..629L}
{Lai}, D. 2001, Reviews of Modern Physics, 73, 629,
  \dodoi{10.1103/RevModPhys.73.629}

\bibitem[{{Lo} {et~al.}(2013){Lo}, {Miller}, {Bhattacharyya}, \&
  {Lamb}}]{2013ApJ...776...19L}
{Lo}, K.~H., {Miller}, M.~C., {Bhattacharyya}, S., \& {Lamb}, F.~K. 2013, \apj,
  776, 19, \dodoi{10.1088/0004-637X/776/1/19}

\bibitem[{{Miller} \& {Lamb}(2015)}]{2015ApJ...808...31M}
{Miller}, M.~C., \& {Lamb}, F.~K. 2015, \apj, 808, 31,
  \dodoi{10.1088/0004-637X/808/1/31}

\bibitem[{{Miller} {et~al.}(2019){Miller}, {Lamb}, {Dittmann}, {Bogdanov},
  {Arzoumanian}, {Gendreau}, {Guillot}, {Harding}, {Ho}, {Lattimer}, {Ludlam},
  {Mahmoodifar}, {Morsink}, {Ray}, {Strohmayer}, {Wood}, {Enoto}, {Foster},
  {Okajima}, {Prigozhin}, \& {Soong}}]{Miller2019}
{Miller}, M.~C., {Lamb}, F.~K., {Dittmann}, A.~J., {et~al.} 2019, \apjl, 887,
  L24, \dodoi{10.3847/2041-8213/ab50c5}

\bibitem[{{Payne} \& {Melatos}(2004)}]{2004MNRAS.351..569P}
{Payne}, D.~J.~B., \& {Melatos}, A. 2004, \mnras, 351, 569,
  \dodoi{10.1111/j.1365-2966.2004.07798.x}

\bibitem[{{P{\'e}tri}(2015)}]{2015MNRAS.450..714P}
{P{\'e}tri}, J. 2015, \mnras, 450, 714, \dodoi{10.1093/mnras/stv598}

\bibitem[{{P{\'e}tri}(2016)}]{2016MNRAS.463.1240P}
---. 2016, \mnras, 463, 1240, \dodoi{10.1093/mnras/stw2050}

\bibitem[{{P{\'e}tri}(2020)}]{2020MNRAS.tmp.3024P}
---. 2020, \mnras, \dodoi{10.1093/mnras/staa3086}

\bibitem[{{Philippov} \& {Spitkovsky}(2018)}]{Philippov2018}
{Philippov}, A.~A., \& {Spitkovsky}, A. 2018, \apj, 855, 94,
  \dodoi{10.3847/1538-4357/aaabbc}

\bibitem[{{Ploeg} {et~al.}(2020){Ploeg}, {Gordon}, {Crocker}, \&
  {Macias}}]{2020arXiv200810821P}
{Ploeg}, H., {Gordon}, C., {Crocker}, R., \& {Macias}, O. 2020, arXiv e-prints,
  arXiv:2008.10821.
\newblock \doarXiv{2008.10821}

\bibitem[{{Potekhin} {et~al.}(2014){Potekhin}, {Chabrier}, \&
  {Ho}}]{2014A&A...572A..69P}
{Potekhin}, A.~Y., {Chabrier}, G., \& {Ho}, W.~C.~G. 2014, \aap, 572, A69,
  \dodoi{10.1051/0004-6361/201424619}

\bibitem[{{Psaltis} \& {Johannsen}(2012)}]{2012ApJ...745....1P}
{Psaltis}, D., \& {Johannsen}, T. 2012, \apj, 745, 1,
  \dodoi{10.1088/0004-637X/745/1/1}

\bibitem[{{Psaltis} \& {{\"O}zel}(2014)}]{2014ApJ...792...87P}
{Psaltis}, D., \& {{\"O}zel}, F. 2014, \apj, 792, 87,
  \dodoi{10.1088/0004-637X/792/2/87}

\bibitem[{{Riley} {et~al.}(2019){Riley}, {Watts}, {Bogdanov}, {Ray}, {Ludlam},
  {Guillot}, {Arzoumanian}, {Baker}, {Bilous}, {Chakrabarty}, {Gendreau},
  {Harding}, {Ho}, {Lattimer}, {Morsink}, \& {Strohmayer}}]{Riley2019}
{Riley}, T.~E., {Watts}, A.~L., {Bogdanov}, S., {et~al.} 2019, \apjl, 887, L21,
  \dodoi{10.3847/2041-8213/ab481c}

\bibitem[{{Ruderman}(1991{\natexlab{a}})}]{1991ApJ...366..261R}
{Ruderman}, M. 1991{\natexlab{a}}, \apj, 366, 261, \dodoi{10.1086/169558}

\bibitem[{{Ruderman}(1991{\natexlab{b}})}]{1991ApJ...382..576R}
{Ruderman}, R. 1991{\natexlab{b}}, \apj, 382, 576, \dodoi{10.1086/170744}

\bibitem[{{Salmi} {et~al.}(2020){Salmi}, {Suleimanov}, {N{\"a}ttil{\"a}}, \&
  {Poutanen}}]{2020A&A...641A..15S}
{Salmi}, T., {Suleimanov}, V.~F., {N{\"a}ttil{\"a}}, J., \& {Poutanen}, J.
  2020, \aap, 641, A15, \dodoi{10.1051/0004-6361/202037824}

\bibitem[{{Spitkovsky}(2006)}]{2006ApJ...648L..51S}
{Spitkovsky}, A. 2006, \apjl, 648, L51, \dodoi{10.1086/507518}

\bibitem[{{Timokhin}(2006)}]{2006MNRAS.368.1055T}
{Timokhin}, A.~N. 2006, \mnras, 368, 1055,
  \dodoi{10.1111/j.1365-2966.2006.10192.x}

\bibitem[{{Timokhin} \& {Arons}(2013)}]{2013MNRAS.429...20T}
{Timokhin}, A.~N., \& {Arons}, J. 2013, \mnras, 429, 20,
  \dodoi{10.1093/mnras/sts298}

\bibitem[{{Timokhin} \& {Harding}(2015)}]{2015ApJ...810..144T}
{Timokhin}, A.~N., \& {Harding}, A.~K. 2015, \apj, 810, 144,
  \dodoi{10.1088/0004-637X/810/2/144}

\bibitem[{{Venter} {et~al.}(2015){Venter}, {Kopp}, {Harding}, {Gonthier}, \&
  {B{\"u}sching}}]{2015ApJ...807..130V}
{Venter}, C., {Kopp}, A., {Harding}, A.~K., {Gonthier}, P.~L., \&
  {B{\"u}sching}, I. 2015, \apj, 807, 130, \dodoi{10.1088/0004-637X/807/2/130}

\bibitem[{{Vigan{\`o}} {et~al.}(2013){Vigan{\`o}}, {Rea}, {Pons}, {Perna},
  {Aguilera}, \& {Miralles}}]{2013Vigano}
{Vigan{\`o}}, D., {Rea}, N., {Pons}, J.~A., {et~al.} 2013, \mnras, 434, 123,
  \dodoi{10.1093/mnras/stt1008}

\bibitem[{{Wasserman} \& {Shapiro}(1983)}]{1983ApJ...265.1036W}
{Wasserman}, I., \& {Shapiro}, S.~L. 1983, \apj, 265, 1036,
  \dodoi{10.1086/160745}

\end{thebibliography}
\end{document}